\documentclass[12pt]{article}

\usepackage{latexsym}
\usepackage{bbm}
\usepackage{amsopn}
\usepackage{amsfonts}
\usepackage{amssymb}
\usepackage{amsthm}
\usepackage{amsmath}

\usepackage{epsfig,graphics}
\usepackage{t1enc}
\usepackage[cp1250]{inputenc}
\usepackage{a4wide}

\renewcommand{\theequation}{\arabic{section}.\arabic{equation}}

\newtheorem{Theorem}{Theorem}[section]
\newtheorem{Proposition}[Theorem]{Proposition}
\newtheorem{Lemma}[Theorem]{Lemma}
\newtheorem{Corollary}[Theorem]{Corollary}

\newcommand{\xxfull}{x_1,x_2,x_3,x_4,x_5,x_6,x_7,x_8}

\newcommand{\xx}{x_1,\ldots,x_8}
\newcommand{\xxx}{x_1,\ldots,x_9}

\renewcommand{\proof}{{\noindent\bf Proof: }}
\newcommand{\koniec}{\hfill$\Box$}
\newcommand{\macierz}[4]
{
    \left[\begin{array}{c|c}
    #1&#2\\
    \hline \rule{0pt}{3ex}#3&#4
    \end{array}\right]
}
\newcommand{\e}{\,\mathrm{e}}
\newcommand{\tr}{\mathrm{tr}}
\newcommand{\id}{{\mathbbm 1}}
\newcommand{\re}{\mathrm{Re}}
\newcommand{\im}{\mathrm{Im}}
\newcommand{\D}{\mathbf{D}}
\newcommand{\R}{\mathbb{R}}
\newcommand{\C}{\mathbb{C}}
\newcommand{\Z}{\mathbb{Z}}
\newcommand{\diag}{{\rm diag}}
\newcommand{\mc}{\mathcal}
\newcommand{\mb}{\mathbf}

\newcommand{\old}[1]{}
\newcommand{\maxtor}{T}

\begin{document}

\title{\bf On the Stratified Classical Configuration Space of Lattice QCD}

\author{
    S.~Charzy\'nski and J.~Kijowski\\
    Center for Theoretical  Physics, Polish Academy of Sciences\\
    al. Lotnik\'ow 32/46, 02-668 Warsaw, Poland\\
    \ \\
    G.~Rudolph and M.~Schmidt\\
    Institut f\"ur Theoretische Physik, Universit\"at Leipzig\\
    Augustusplatz 10/11, 04109 Leipzig, Germany\\
    }

\maketitle

\\


%%%%%%%%%%%%%%%%%%%%%%%%%%%%%%%%%%%%%%%%%%%%%%%%%%%%%%%%%%%%%%%%%%%%%%%%%%%%%%%%%%%%%%%%%%%

\begin{abstract}
The stratified structure of the configuration space $\mb G^N = G
\times \dots \times G$  reduced with respect to the action of $G$
by inner automorphisms is investigated for $G = SU(3) \, .$ This
is a finite dimensional model coming from lattice QCD. First, the
stratification is characterized algebraically, for arbitrary $N$.
Next, the full algebra of invariants is discussed for the cases $N
= 1$ and $N =2 \, .$  Finally, for $N = 1$ and $N =2 \, ,$ the
stratified structure is investigated in some detail, both in terms
of invariants and relations and in more geometric terms. Moreover,
the strata are characterized explicitly using local cross
sections.

\end{abstract}

\newpage

\newpage

{\tableofcontents}

\newpage
%%%%%%%%%%%%%%%%%%%%%%%%%%%%%%%%%%%%%%%%%%%%%%%%%%%%%%%%%%%%%%%%%%%%%%%%%%%%%%%%%%%%%%%%%%%%%
\setcounter{equation}{0}
\section{Introduction}
\label{Introduction}
%%%%%%%%%%%%%%%%%%%%%%%%%%%%%%%%%%%%%%%%%%%%%%%%%%%%%%%%%%%%%%%%%%%%%%%%%%%%%%%%%%%%%%%%%%%%%

If one wants to analyze the
non-perturbative structure of gauge theories, one should start with
clarifying basic structures like that of the field algebra, the observable algebra
and the superselection structure of the Hilbert space of physical
states. It is clear that the standard Doplicher-Haag-Roberts
theory \cite{DHR,DR} for models, which do not contain
massless particles, does not apply here. Nonetheless, there are
interesting partial results within the framework of general
quantum field theory both for quantum electrodynamics (QED) and
for non-abelian models, see \cite{Bu,JF,SW1,S}.

To be rigorous, one can put the system on a finite lattice, leaving the
(extremely complicated task) of constructing the full continuum limit,
for the time being, aside. This way, one gets rid of complicated functional
analytical problems, but the gauge theoretical problems one is interested in are still
present within this setting. For basic notions concerning lattice gauge theories
(including fermions) we refer to \cite{Seiler} and references therein.
Our approach is Hamiltonian, thus, we put the model on a finite (regular) cubic lattice.
In this context, we have formulated (and in the
meantime partially solved \cite{KRT,KRS,KRS1,KR,KR1})
the following programme:

\begin{enumerate}
\item
Describe the field algebra ${\mathfrak A}_{\Lambda}$ in terms of generators and defining
relations and endow it with an appropriate functional analytical structure
\item
Describe the observable algebra ${\mathfrak O}_{\Lambda}$ (algebra of gauge invariant operators, fulfilling the
Gauss law) in terms of generators and relations
\item
Analyze the mathematical structure of ${\mathfrak O}_{\Lambda}$ and endow it with an
appropriate functional analytical structure
\item
Classify all irreducible representations of ${\mathfrak O}_{\Lambda}$
\item
Investigate dynamics in terms of observables
\end{enumerate}

Finally, of course, one wants to
construct the continuum limit. As already mentioned, in full generality, this is 
an extremely complicated problem of constructive field theory.
However, the results obtained until now suggest that there is some
hope to control the thermodynamical limit, see \cite{KRT} for a
heuristic discussion. We also mention that for simple toy models,
these problems can be solved, see \cite{FM}.

In \cite{KR1} we have started to investigate the structure of the
field and the observable algebra of lattice QCD. In these papers we took the
attitude of implementing the constraints on the quantum level.
It is well known that there is another possibility: First, one reduces the
classical phase space and then one formulates the quantum theory over this
reduced phase space. Since the action of the gauge group can have several orbit
types, the first step has to be done using singular Marsden-Weinstein
reduction \cite{L--S}. Then the reduced phase space has
the structure of a stratified symplectic space. Quantization procedures for
such spaces have been worked out recently or are still under investigation
\cite{Hue}. As an important ingredient for both reduction and quantization, in
this paper, we study the stratified structure of the reduced classical
configuration space. For QCD on a finite lattice, this is given by
the orbit space of the
action of $SU(3)$ on $SU(3)^N = SU(3)\times\cdots\times SU(3)$ by inner
automorphisms.

Our paper is organized as follows: In Section \ref{Basics} we give
a precise formulation of the problem and we discuss the basic
tools used in this paper. In Section \ref{stratification}, the
stratification of the reduced configuration space is characterized
algebraically for arbitrary $N$. Next, in Section
\ref{invariantsorbspace} the full algebra of invariants is
discussed for the cases $N = 1$ and $N =2 \, .$ Finally, in
Sections \ref{N=1} and \ref{N=2}  the stratified structure is
investigated for $N = 1$ and $N =2$ in some detail, both in terms
of invariants and relations and in more geometric terms. Moreover,
the strata are characterized explicitly using local cross
sections.

%%%%%%%%%%%%%%%%%%%%%%%%%%%%%%%%%%%%%%%%%%%%%%%%%%%%%%%%%%%%%%%%%%%%%%%%

\setcounter{equation}{0}
\section{Basics}
\label{Basics}

%%%%%%%%%%%%%%%%%%%%%%%%%%%%%%%%%%%%%%%%%%%%%%%%%%%%%%%%%%%%%%%%%%%%%%%%

\noindent
We consider QCD on a finite regular cubic lattice $\Lambda$
in the Hamiltonian framework. In this context, the classical gluonic potential is
approximated by its parallel transporter:
$$
{\Lambda}^1 \ni (x,y) \rightarrow g_{(x,y)} \in G \, ,
$$
where $G = SU(3)$ and ${\Lambda}^1$ denotes the set of $1$-dimensional
elements (links) of $\Lambda\, .$
Thus, the classical configuration space ${\cal C}_{(x,y)}$ over a given link $(x,y)$ is isomorphic
to the group manifold $G$ and the classical phase space over $(x,y)$ is isomorphic to
$$
\quad T^* G \cong {\mathfrak g}^* \times G \ .
$$
Thus, the (gluonic) lattice configuration space is given by
 \begin{equation}
 \label{lattice-confispace}
{\cal C}_{\Lambda} = \prod_{(x,y) \in \Lambda^1 } {\cal C}_{(x,y)} \, .
 \end{equation}
It is obviously isomorphic to the product
$$
\mb G^L := \underbrace{G \times \cdots \times G}_{L}~,
$$
with $L$ denoting the number of lattice links. The corresponding
phase space is a product of phase spaces of the above type.
Gauge transformations act on parallel transporters by
$$
\quad g_{(x,y)} \mapsto g_{(x,y)}^{\prime} = g_x \cdot g_{(x,y)} \cdot
g_y^{-1} \, ,
$$
with
$$
\Lambda^0 \ni x \mapsto g_x \in G
$$
and $\Lambda^0$ denoting the set of $0$-dimensional elements (sites) of
$\Lambda \, .$ These transformations
induce transformations of the phase space over $(x,y)\, .$
Thus, the lattice gauge group is given by
\begin{equation}
  \label{lattice--gaugegroup}
G_{\Lambda} = \prod_{x
  \in \Lambda^0 } G_x \, ,
\end{equation}
with every $G_x$ being a copy of $G \, .$

The above symmetry can be easily reduced using the following technique:
We choose a {\em lattice tree}, which consists of a fixed lattice point (root)
$x_0$ and a subset of ${\Lambda}^1$ such that for every lattice site $x$
there is a unique lattice path from $x$ to $x_0 \, .$
Now, we can fix the gauge on every on--tree link and we can
parallel transport every off-tree configuration variable
to the point $x_0$.
This can be viewed as a reduction with respect to the pointed lattice
gauge group
\begin{equation}
  \label{pointed--gaugegroup}
G^0_{\Lambda} = \prod_{x_0 \neq x
  \in \Lambda^0 } G_x \, .
\end{equation}
We end up with a partially reduced configuration space being
isomorphic to $\mb G^N$, with $N$ denoting the number of off-tree
links. The corresponding phase space is given by the cotangent bundle $T^\ast\mb
G^N$.
The reduced gauge group is $G_{x_0}\equiv G$, acting via inner
automorphisms $G \ni g \mapsto  Ad_g \in Aut(\mb
G^N)$:
$$
Ad_g(g_1,\ldots,g_N) =(g\cdot g_1\cdot g^{-1},g\cdot
g_2\cdot g^{-1},\ldots,g\cdot g_N\cdot g^{-1}) \, .
$$

Thus, we have a finite dimensional Hamiltonian system with 
symmetry group $G$. Since this action has
several orbit types, quantization turns out to be a complicated
task. Usually, the non-generic strata occuring here are omitted.
If one wants to include them consistently, one has to develop
a quantum theory over a stratified set. One option to do this is
to perform quantization after reduction, i.e., to quantize the
reduced phase space of $\mb G^N$. This is a stratified symplectic
space which is constructed from $T^\ast \mb G^N$ by singular
Marsden-Weinstein reduction \cite{L--S}. By properly implementing
the tree gauge on the level of the phase space, it can be shown
that this space is isomorphic, as a stratified symplectic space,
to the reduced phase space of the full lattice gauge theory
\cite{Schm}. This completely justifies the use of the tree gauge
in this approach. The reduced phase space of $\mb G^N$ is a bundle
over the reduced configuration space
 \begin{equation}
 \label{red--configspace}
\hat {\cal C}_{\Lambda} \cong \mb G^N/Ad_G~.
 \end{equation}
In this work, we investigate $\hat {\cal C}_{\Lambda}$ for $N=1$
and $2$.

Our strategy is as follows:\\
i)~It is well-known that orbit types of the action of a Lie group $G$
on a manifold $M$ are classified by conjugacy classes of
stabilizers $[G_m]$, $m\in M$, of the group action. Moreover, the
orbit of an element $m$ is diffeomorphic to $G/G_m$. Thus, in Section
\ref{stratification}, we list the orbit types by calculating their
stabilizers. This is done for {\em arbitrary} $N \, .$
Moreover, all orbit types will be characterized
algebraically, in terms of properties of eigenvectors and eigenvalues
of representatives. \\
ii)~Next, in order to investigate the geometric structure of $\hat {\cal
C}_{\Lambda}$, we make use of basic facts from invariant
theory. According to  \cite{schwarz},
if we have an action of a Lie group $G$ on a
manifold $M$ with a finite number of orbit types, then the orbit space
of this action can be characterized as follows:
Let $(\rho_1\ldots\rho_p)$ be a set of generators of the algebra of invariant
polynomials of the $G$-action on $M\,.$ They define a mapping
$$
\rho=(\rho_1\ldots\rho_p): M \longrightarrow \mathbb{R}^p \, ,
$$
which induces a homeomorphism of the orbit space $X:= M/G$ onto
the image of $\rho$ in $\mathbb{R}^p \, .$ Next, restricting our
attention to the case of $G$ being an $(n \times n)$-matrix group
and $M = \mb G^N\, ,$ we can use general results as developed in
\cite{procesi}: The algebra of polynomials, which are invariant
under simultaneous conjugation of $N$ matrices is generated by
traces of products of these matrices,
 \begin{equation}
 \label{invariantfamily}
\mb G^N \ni (g_1,\ldots,g_N) \mapsto \tr\left(g_{i_1} g_{i_2}
\cdots g_{i_k}\right) \in {\mathbb C} \, ,
 \end{equation}
with $k \leqslant 2^n-1 \, .$
Moreover, for $Gl(n,\mathbb{R})$, the full set of relations
between generators is given by the so--called fundamental trace identity
 \begin{equation}
 \label{fti}
\sum_{\sigma \in S_{n+1}} \mathrm{sgn}(\sigma)\cdot\prod_{(i_1,\ldots,i_j)}
\tr(g_{i_1}\cdots g_{i_j})=0 \, ,
\end{equation}
where $(i_1,\ldots,i_j)$ ranges over the set of all cycles
of the cycle decomposition of the permutation $\sigma\, .$
In the case under consideration, $G = SU(n) \, ,$ we have two additional relations induced
from the two invariant tensors of $SU(n)$, see \cite{Weyl},
\begin{eqnarray}
\label{tr--id1} \tr(g g^{\dagger}) & = & n \, ,\\
\label{tr--id2} det(g) & = &1 \, .
\end{eqnarray}
Relations (\ref{tr--id1}) and  (\ref{tr--id2}) imply the following
form of the characteristic polynomial of
$g \in G = SU(3)$:
 \begin{equation}
 \label{charpol}
\chi_g(\lambda)=\lambda^3-\tr(g)\lambda^2+\overline{\tr(g)}\lambda-1 \, .
 \end{equation}
The above listed facts enable us to characterize the configuration
space in terms of invariant generators and relations. First, in
Section \ref{invariantsorbspace}, we investigate the algebra of
invariants and their relations. Next, in Section \ref{N=1} and
Subsection \ref{Invariants and Relations} we study the mapping
$\rho$ in some detail. For $N=1$ we solve the problem completely,
that means we find the range of $\rho$ and characterize $\hat
{\cal C}_{\Lambda}$ as a compact subset of ${\mathbb R}^2 \, .$
For $N=2$, we will find a unique characterization of each orbit
type in terms of invariants. But to find the range of $\rho$,
defined in terms of a number of {\em inequalities} between
invariants, turns out to be a complicated problem. Therefore, this
will be discussed in a separate paper, see \cite{ChKRS}. There, we
will present a complete topological characterization of $\hat
{\cal C}_{\Lambda}$ for $N=2$ as a CW-complex. \\
iii)~We present a somewhat detailed
geometric characterization of all occuring strata in terms of subsets and
quotients of $SU(3)$, see Subsection \ref{N=2--geomstr}.\\
iv)~Using a principal bundle atlas of $SU(3)\, ,$ viewed as an
$SU(2)$-bundle over $S^5\, ,$ we construct representatives of orbits for all occuring strata,
see Subsection \ref{Representatives}.

%%%%%%%%%%%%%%%%%%%%%%%%%%%%%%%%%%%%%%%%%%%%%%%%%%%%%%%%%%%%%%%%%%%%%%%%%%%%%%%%%%%%%%%%%

\setcounter{equation}{0}
\section{The Stratification of the Configuration Space}
\label{stratification}

%%%%%%%%%%%%%%%%%%%%%%%%%%%%%%%%%%%%%%%%%%%%%%%%%%%%%%%%%%%%%%%%%%%%%%%%%%%%%%%%%%%%%%%%%

First, let
us consider the case $N=1$.

\begin{Theorem}\label{stratN=1}

The adjoint action of $SU(3)$ on $\mb G^1 \equiv SU(3)$ has three
orbit types, corresponding to three conjugacy classes of
stabilizers of dimension $2$, $4$ and $8$, respectively. The orbit
space $\mb G^1/Ad_{SU(3)}$ decomposes into three strata
characterized by the following conditions:

\begin{enumerate}
\item
If $g$ has three different eigenvalues then its
stabilizer is $U(1)\times U(1)$ and $g$ belongs to the generic
stratum.

\item If $g$ has two different eigenvalues then its
stabilizer is $U(2)$.

\item If $g$ has only one eigenvalue then it belongs to the centre $\mc Z$ and 
its stabilizer is $G=SU(3)$.
\end{enumerate}

\end{Theorem}

\proof
Up to conjugacy, we may assume that $g =
\diag(\lambda_1,\lambda_2,\lambda_3)$. In case 1, the $\lambda_i$
are pairwise distinct. Hence, the stabilizer of $g$ is
 \begin{equation}
 \label{stabilizer1}
H_g =  \left\{
\diag(\alpha,\beta,\gamma)
 \, | \,
\alpha, \beta , \gamma \in U(1)
 \, , \,
\alpha \cdot \beta \cdot \gamma = 1
 \right\}
 \cong
U(1) \times U(1) \, .
\end{equation}
In case 2, up to conjugacy, $\lambda_1\neq\lambda_2 = \lambda_3$.
Then the stabilizer of $g$ is
 \begin{equation}
 \label{stabilizer2}
H_g = \left\{
 \left. \macierz{(\det V)^{-1}}{}{}{V} \, \right| \, V \in U(2)
 \right\}
 \cong
U(2) \, .
\end{equation}
In case 3, $\lambda_1=\lambda_2=\lambda_3$, i.e., $g$ is a
multiple of the identity. Hence, its stabilizer is $G=SU(3)$.
Finally, it is clear that cases 1--3 exhaust all possible values
of the $\lambda_i$.
 \qed

Next, we deal with the general case.

\begin{Theorem}\label{stratyfikacja}

The adjoint action of $SU(3)$ on $\mb G^N$, $N\geqslant 2$, has
five orbit types, corresponding to five conjugacy classes of
stabilizers of dimension $0,1,2,4$ and $8$, respectively. The
orbit space $\mb G^N/Ad_{SU(3)}$ decomposes into five strata
characterized by the following conditions. Denote $\mb g :=
(g_1,\ldots,g_N)$.

\begin{enumerate}
\item If $g_1,\ldots,g_N$ have no common eigenspace then the
stabilizer of $\mb g$ is $H_{\mb g} = \mc Z$ and $\mathbf g$
belongs to the generic stratum.

\item
If $g_1,\ldots,g_N$ have exactly one common
$1$-dimensional eigenspace then $H_{\mb g} \cong U(1)$.

\item
If $g_1,\ldots,g_N$ have three (different) common ($1$-dimensional)
eigenspaces then $H_{\mb g} \cong U(1)\times U(1)$.

\item
If $g_1,\ldots,g_N$ have a common
$2$-dimensional eigenspace then $H_{\mb g} \cong U(2)$.

\item
If $g_1,\ldots,g_N$ have a $3$-dimensional
common eigenspace, i.e., if they all are proportional to the
identity then $H_{\mb g} = G = SU(3)$.
\end{enumerate}

\end{Theorem}

\proof

If there are two eigenvectors $e_1$ and $e_2$, common for all
matrices $g_1,\ldots,g_N$, then also their vector product $e_1
\times  e_2$ is a common eigenvector. If $e_1$ and $e_2$ are not
orthogonal, then the 2-dimensional space ${\bf P}$ spanned by them
is a common eigenspace. This means that the pair $(e_1,e_2)$ can
be replaced by any orthonormal basis of ${\bf P}$. This implies
that if ${\mathbf g}$ is not of type 1 or 2, its elements can be
jointly diagonalized. We conclude that the above types exhaust all
possible cases.
\\
Next we calculate the stabilizer for each case.

\begin{enumerate}

\item Assume that the stabilizer of $\mb g$ contains an
element $s\not\in\mc Z$. Then $s$ has at least 2 different
eigenvalues. One of these must be nondegenerate. Since the
corresponding eigenspace is left invariant by all $g_i$ and since
it is 1-dimensional, it is an eigenspace of all $g_i$, in
contradiction to the assumption.

\item Since the $g_i$ have a common eigenvector $e_1$, up to
conjugacy, we may assume that
$$
g_i = \macierz{a_i}{0}{0}{B_i}\,,
$$
where $B_i\in U(2)$. Then $H_{\mb g}$ contains the subgroup
 \begin{equation}
 \label{Gclfot2}
 \left
\{\left. \macierz{\alpha}{0}{0}{\beta {\bf 1}}
 \,\right|\,
\alpha,\beta\in U(1) \, , \, \beta^2 = \overline{\alpha}
 \right\}
 \cong
U(1)\,.
 \end{equation}
Conversely, let $s\in H_{\mb g}$. Since the common eigenspace of
the $g_i$ is $1$-dimensional, $e_1$ is also an eigenvector of $s$.
Then
$$
  s=\macierz{\alpha}{0}{0}{A} \,,
$$
where $A\in U(2)$. Again up to conjugacy, we may assume that $A =
\diag(\beta,\gamma)$. If $\beta \neq \gamma$ then the $B_i$ must
also be diagonal, because they commute with $A$. Then the $g_i$
have more than one common eigenspace, which contradicts the
assumption. Hence $\beta = \gamma$ and $H_{\mb g}$ coincides with
the subgroup \eqref{Gclfot2}.

\item Choose a basis in ${\mathbb C}^3$, which jointly
diagonalizes all the matrices $g_1,\ldots,g_N$,
$$
 g_i= \left[
 \begin{array}{ccc}
 a_i&0&0\\
 0&b_i&0\\
 0&0&c_i
 \end{array}
\right] \ .
$$
The non-existence of a $2$-dimensional eigenspace means that none among the
three equations $a_i = b_i$, $b_i = c_i$ and
$c_i = a_i$, is fulfilled for all $i$. This implies that
any matrix which commutes with all matrices $g_1,\ldots,g_N$ must be diagonal, too.
Whence, the stabilizer $H_{\mathbf g}$  is of the form
(\ref{stabilizer1}).

\item The orthogonal complement of the 2-dimensional common
eigenspace of the $g_i$ is a one-dimensional common eigenspace.
Thus, up to conjugacy,
$$
g_i=\macierz{a_i}{0}{0}{b_i {\bf 1}}
$$
and $H_{\mb g}$ contains the subgroup (\ref{stabilizer2}).
Conversely, let $s\in H_{\mb g}$. The non-existence of a
$3$-dimensional eigenspace means that there is $i_0$ such that
$a_{i_0}\neq b_{i_0}$. Then
$$
s=\macierz{(\det V)^{-1}}{0}{0}{V} \ ,
$$
with $V \in U(2)$.  Whence, $H_{\mathbf g}$ coincides with the
subgroup (\ref{stabilizer2}).

\item In this case, all matrices $g_1,\ldots,g_N$ belong to $\mathcal Z$, so
the statement is obvious.\qed
\end{enumerate}

Observe that types 1 and 3 may be uniquely characterized as
follows:

\begin{Corollary}
\label{pairsandtriples}
\noindent
\begin{enumerate}
\item
The matrices $g_1,\ldots,g_N$ have no common eigenvector if
and only if there exists a pair $(g_i,g_j)$ or a triple
$(g_i,g_j,g_k)$ of elements not possessing any common eigenvector.

\item
Suppose that $g_1,\ldots,g_N$ have three (different) common
(1-dimensional) eigenspaces. There does not exist a common $2$-dimensional
eigenspace if and only if there exists an element $g_i$ with three
different eigenvalues or a pair $(g_i,g_j)$ such that each of its
elements has exactly two different eigenvalues and non-degenerate
eigenvalues correspond to different eigenvectors.

\end{enumerate}

\end{Corollary}

\proof

\noindent 1. If there exists a pair $(g_i,g_j)$ or a triple
$(g_i,g_j,g_k)$ having no common eigenvector then, obviously,
there is no common eigenvector for all of them. Conversely, assume 
that every triple $(g_i,g_j,g_k)$ has
a common eigenvector. We prove that in this case there exists a
common eigenvector for all matrices $g_1,\ldots,g_N$. First,
observe that it is sufficient to consider the case when none of
the matrices $g_1,\ldots,g_N$ is fully degenerate (i.e.~$g_i
\notin \mathcal Z$). This means that every $g_i$ has at least two
different eigenvalues.

The proof goes via induction: for $K \ge
3$ we show that if any subset of $\mb g$ of $K$ elements has a
common eigenvector, then the same is true for any subset of $K+1$ elements. 
Thus, take a
subset $(g_1,\ldots,g_{K+1})$. For each $i=1,\dots,K+1$, skip
$g_i$ and choose a common eigenvector $v_i$ of the remaining set
of $K$ elements. If there exist $i\neq j$ such that $v_i$ and
$v_j$ are parallel then they both are common eigenvectors of
$g_1,\ldots,g_{K+1}$. Otherwise, there exist $i\neq j$ such that
$v_i$ and $v_j$ are not orthogonal, because there cannot be more
than 3 mutually orthogonal vectors in ${\mathbb C}^3$. Suppose
that $v_K$ and $v_{K+1}$ is such a pair. It spans a
$2$-dimensional subspace ${\bf P}\subset {\mathbb C}^3$. Since
$v_K$, $v_{K+1}$ are common, non-orthogonal eigenvectors of
$g_1,\ldots,g_{K-1}$, ${\bf P}$ is a common eigenspace of these
elements. Now consider $v_1$. Since it is an eigenvector of $g_2$
and since, by assumption, $g_2$ is not proportional to the
identity, $v_1$ must either belong to ${\bf P} $ or be orthogonal
to ${\bf P} $. But in both cases it is also an eigenvector of
$g_1$ and, therefore, a common eigenvector of
$g_1,\ldots,g_{K+1}$.

\noindent
2. In this case all matrices $g_1,\ldots,g_N$ can be
jointly diagonalized. If one of them has 3 different eigenvalues
(i.e., it has no $2$-dimensional eigenspace), then there is no common
$2$-dimensional eigenspace ${\mb P} $ for all of them. Suppose
that this is not the case, i.e., that every $g_i$ has a
$2$-dimensional eigenspace ${\mb P}_i$. There will be no {\em common}
$2$-dimensional eigenspace if and only if there exist $i,j$ such that
${\mb P}_i \ne {\mb P}_j$. Then also the non-degenerate eigenspaces $\mb Q_i$
and $\mb Q_j$ of $g_i$ and $g_j$ do not coincide, because they are given by the orthogonal 
complements of $\mb P_i$ and $\mb P_j$, respectively. Hence, the decomposition
of $\C^3$ into common eigenspaces of $g_i$ and $g_j$ is $\mb Q_i\oplus\mb Q_j
\oplus \mb P_i\cap\mb P_j$.
\qed

%%%%%%%%%%%%%%%%%%%%%%%%%%%%%%%%%%%%%%%%%%%%%%%%%%%%%%%%%%%%%%%%%%%%%%%%%%%%%%%%%%%%%%%%%
\setcounter{equation}{0}
\section{The Algebra of Invariants}
\label{invariantsorbspace}
%%%%%%%%%%%%%%%%%%%%%%%%%%%%%%%%%%%%%%%%%%%%%%%%%%%%%%%%%%%%%%%%%%%%%%%%%%%%%%%%%%%%%%%%%

In this section, we analyze the algebra of invariants for
$N=1$ and $N=2 \, .$
We start with invariant monomials built from one matrix.
\begin{Lemma}
\label{indepinv}
\noindent
The invariants $\tr(g^i)$ can be uniquely expressed in terms of
$\tr(g)\, , $ for any integer $i \, .$
\end{Lemma}
\proof Recall formula \ref{charpol} for the characteristic polynomial of $g \in SU(3):$
$$
\chi_g(\lambda)=\lambda^3-\tr(g)\lambda^2+\overline{\tr(g)}\lambda-1.
$$
Thus, by the Cayley-Hamilton theorem, we have
 \begin{equation}
 \label{cayl1}
g^3 - \tr(g)g^2 + \overline{\tr(g)}g-\id=0 \, .
 \end{equation}
Multiplying both sides of (\ref{cayl1}) by $g^{-1}$ we obtain:
 \begin{equation}
 \label{cayl2}
g^2-\tr(g)g+\overline{\tr(g)}-g^{-1}=0.
\end{equation}
Taking the trace of both sides we get
 \begin{equation}
 \label{kwadrat}
\tr(g^2)=\left(\tr(g)\right)^2-2\overline{\tr(g)}.
 \end{equation}
Analogously, multiplying (\ref{cayl1}) by $g^i$, $i\geqslant 1$ and
taking the trace one gets
formulae for $\tr(g^{i+2})$ in terms of traces of
$\tr(g^{i+1})$, $\tr(g^i)$ and $\tr(g)$. So by induction $\tr(g^i)$
is uniquely given by $\tr(g)$.
For negative $i\, ,$ the statement now follows from \ref{tr--id1}.
\qed

So in case $N=1\, ,$ the  algebra of invariant functions has only one
generator. The case $N=2$ is more complicated. Its characterization
in terms of invariant generators will be given in Theorem
\ref{alginvN=2}.

\begin{Lemma}
\label{indepinv3}
The invariants $\tr(g^ih^j)$ can be uniquely expressed in terms of
the following set of independent invariants:
 \begin{equation}
 \label{indepinv2}
\left\{\tr(g), \tr(h), \tr(gh), \tr(g^2h)\right\} \, .
 \end{equation}
\end{Lemma}
\proof
First,
substituting $g\rightarrow gh$ in (\ref{cayl2}) and
multiplying both sides by $g^{-1}$ to the left we get:
 \begin{equation}
 \label{fundrel}
hgh-\tr(gh)h+\overline{\tr(gh)}g^{-1}-(ghg)^{-1}=0.
 \end{equation}
Taking the trace of both sides yields:
 \begin{equation}
 \label{fundrel2}
\tr(gh^2)-\tr(gh)\tr(h)+\overline{\tr(gh)\tr(g)}-\overline{\tr(g^2h)}=0.
 \end{equation}
Thus, from five traces occurring in this equation only four are
independent. In what follows, we express $\tr(gh^2)$ in terms of
the set
$$
\left\{\tr(g), \tr(h), \tr(gh), \tr(g^2h)\right\} \, .
$$
Multiplying (\ref{cayl1}) by $h g^i$ and taking the trace we obtain
 \begin{equation}
 \label{fundrel3}
\tr(hg^{i+3})-\tr(g)\tr(hg^{i+2})+\overline{\tr(g)}\tr(hg^{i+1})
-\tr(hg^i)=0,
 \end{equation}
This equation
enables us to express $\tr(h g^{i+3})$ in terms of $\tr(h g^{i+2})$,
$\tr(h g^{i+1})$ and
$\tr(h g^i)$, so by induction it can be expressed in terms of
$\tr(h g^{2})$,
$\tr(hg)$, $\tr(h)$ and $\tr(g)$.

Starting now from an arbitrary invariant of the form $\tr(g^i h^j)\, ,$
we can use the above procedure recursively. First, we lower the power $i$ of $g$
and then we lower the power $j$ of $h$. We end up with invariants
of the form $\tr(h^m g^l)$, with $k\leqslant 2$, $l\leqslant 2$.
So, to finish the proof it is sufficient to express $\tr(g^2 h^2)$
in terms of the set (\ref{indepinv2}). For that purpose, we use the
fundamental trace identity (\ref{fti}) for $k=4$.
Substituting $g_1=g_2=g$, $g_3=g_4=h$ we obtain:
 \begin{eqnarray}
\tr^2(g)\tr^2(h)-4\tr(hg)\tr(g)\tr(h)-\tr^2(g)\tr(h^2)
-\tr(g^2)\tr^2(h)+2\tr^2(hg)
&&
\\
+\,\,4\tr(g)\tr(h^2g)+\tr(h^2)\tr(g^2)
+4\tr(h)\tr(hg^2)-2\tr(hghg)-4\tr(h^2g^2)&=&0.\nonumber
\end{eqnarray}
Using equation (\ref{kwadrat}) we get
$$
\tr(hghg)=tr\left( (hg)^2 \right)=\tr^2(hg)-2\overline{\tr(hg)}.
$$
This way we obtain a formula for $\tr(h^2g^2)$ in terms of invariants
(\ref{indepinv2}).
\qed

\begin{Lemma}\label{hhgghg}

The invariants $\tr(h^2g^2hg)$ and $\tr(h^2ghg^2)$ have the following
properties:
\begin{enumerate}

\item The sum $\tr(h^2g^2hg)+\tr(h^2ghg^2)$ can be expressed
as a polynomial in invariants of order $k\leqslant 5$,

\item $\re\left(\tr(h^2g^2hg)-\tr(h^2ghg^2)\right)=0$,

\item $\tr(h^2g^2hg)-\tr(h^2ghg^2)=
\frac{1}{3}\tr\left( (hg-gh)^3\right)=\det(hg-gh)$,

\item The invariant $\left(\tr(h^2g^2hg)-\tr(h^2ghg^2)\right)^2$
can be expressed
as a polynomial in the invariants \eqref{indepinv2}
and their complex conjugates.

\end{enumerate}

\end{Lemma}

\proof

\begin{enumerate}

\item Using the fundamental trace identity (\ref{fti}) for $k=4$ and $g_1=hgh$, $g_2=g$,
$g_3=h$, $g_4=g$ we obtain:
 \begin{align}
 \label{symtrid}
2\tr(h^2gh & g^2)+  2\tr(h^2g^2hg)+2\tr(hghghg) 
\\ \nonumber
= ~& 
\tr(h^2g)\tr(g)^2\tr(h)-2\tr(hghg)\tr(g)\tr(h)
-2\tr(h^2g)\tr(g)\tr(hg)
\\ \nonumber
& -
\tr(h^2g)\tr(h)\tr(g^2) -\tr(h^3g)\tr(g)^2+2\tr(hghg)\tr(hg)
+ 4\tr(h^2ghg)\tr(g)
\\ \nonumber
& +
2\tr(h^2g)\tr(hg^2)
+\tr(h^3g)\tr(g^2)+2\tr(hghg^2)\tr(h)\,.
 \end{align}
On the left-hand-side of this equation there are invariants of order
$6$, \ and on the right-hand-side all the invariants are of lower order.
By Lemma \ref{indepinv}, we express $\tr(hghghg)$ as follows
$$
\tr(hghghg)=\tr((hg)^3)=\tr^3(hg)-3\overline{\tr(hg)}tr(hg)+3.
$$
Moving it to the right-hand-side yields the statement.

\item
By substituting $g\rightarrow gh$, $h\rightarrow hg$ in (\ref{fundrel})
we obtain:
$$
\tr(h^2ghg^2)-\tr(h^2g^2)\tr(hg)+\overline{\tr(h^2g^2)\tr(hg)}
-\overline{\tr(h^2g^2hg)}=0.
$$
Taking the real part yields:
$$
\re(\tr(h^2ghg^2))-\re(\tr(h^2g^2hg))=0.
$$

\item

The first equality is obtained by expanding the right-hand-side. The
second one follows from the formula for the determinant of a
$3\times 3$-matrix $A$ in terms of traces,
$$
\det(A)
 =
\frac{1}{3}\tr(A^3) - \frac{1}{2}\tr(A^2)\tr(A) + \frac{1}{6}\tr(A)^3\,.
$$
Nevertheless, it can be checked by direct computation.

\item The explicit formula expressing this invariant in terms of invariants
(\ref{indepinv2}) is lengthy and, therefore, we give it in
Appendix \ref{app_relation}, including some remarks how to derive it.

\end{enumerate}
\qed

\begin{Theorem}\label{alginvN=2} 

Any function on $\mb G^2 = G\times G$ invariant
with respect to the adjoint action of $G$ can be expressed as a polynomial
in the following invariants and their complex conjugates:
\begin{eqnarray}
T_1(g,h)&:=&\tr(g),\nonumber\\
T_2(g,h)&:=&\tr(h),\nonumber\\
T_3(g,h)&:=&\tr(hg),\nonumber\\
T_4(g,h)&:=&\tr(hg^2),\nonumber\\
T_5(g,h)&:=&\tr(h^2g^2hg)-\tr(h^2ghg^2).
\end{eqnarray}
Moreover, for given values of $T_1,\ldots, T_4$, there are at most two 
possible values of $T_5$.

\end{Theorem}

\proof
First we observe that using equation (\ref{cayl2}) we can express
$g^{-1}$ in terms of positive powers of $g$ and $\tr(g)$. This implies
that every invariant can be expressed as a polynomial in traces of products
of only positive powers of matrices $g$ and $h$.

From the general theory \cite{procesi} we know that we can
restrict ourselves to invariants of order $k\leqslant 2^n-1=7$. By
Lemmas \ref{indepinv} and \ref{indepinv3}, all invariants of the
type $\tr(g^k)$, $\tr(h^k)$, $\tr(h^ig^j)$ can be expressed in
terms of $T_1,T_2,T_3,T_4$. Observe that all invariants of order
$k\leqslant 3$ are of this type. In what follows we list
invariants of order $k\leqslant 7$ which are not of this type, and
for each order $k$ we present the method of expressing it in terms
of invariants of lower order and $T_i$.
\begin{itemize}
\item $k=4$: $\tr(hghg)$. By Lemma \ref{indepinv}, we have
$\tr(hghg)=tr((hg)^2)=\tr^2(hg)-2\overline{\tr(hg)}$.
\item $k=5$: $\tr(hghg^2)$, $\tr(h^2ghg)$. Substituting
$h\rightarrow hg$ in (\ref{fundrel2}) we obtain:
$$
\tr(g^2hgh)=\tr(g\cdot hg\cdot hg)=\tr(g\cdot hg)\tr(hg)
-\overline{\tr(g\cdot hg)\tr(g)}+\overline{\tr(g^2\cdot hg)}.
$$
Analogously we deal with  $\tr(h^2ghg)$.
\item $k=6$: $\tr(h^3ghg)$, $\tr(g^3hgh)$, $\tr(h^2g^2hg)$,
$\tr(h^2ghg^2)$, $\tr(hghghg)$.
The invariant $\tr(hghghg)=\tr((hg)^3)$ can be expressed
in terms of $\tr(hg)$ by Lemma \ref{indepinv}.
Next, by Lemma \ref{indepinv3}, we can reduce the power in
$\tr(h^3\cdot ghg)$ and express it in terms of
$\tr(h^2\cdot ghg)$ and other invariants of lower order.
(More precisely, we substitute $h\rightarrow ghg$ into equation
(\ref{fundrel3}) for $i=0$).
We deal with $\tr(g^3hgh)$ analogously. Next, we rewrite $\tr(h^2g^2gh)$
and $\tr(h^2ghg^2)$ in the following way:
\begin{eqnarray*}
\tr(h^2g^2gh)&=&\frac{1}{2}\Big(\tr(h^2g^2gh)+\tr(h^2ghg^2)\Big)
+\frac{1}{2}\Big((\tr(h^2g^2hg)-\tr(h^2ghg^2)\Big)=\\
&=&\frac{1}{2}\Big(\tr(h^2g^2hg)+\tr(h^2ghg^2)\Big)+\frac{1}{2} T_5(g,h)
\, ,\\
\tr(h^2ghg^2)
%&=&\frac{1}{2}\Big(\tr(h^2g^2hg)+\tr(h^2ghg^2)\Big)
%-\frac{1}{2}\Big((\tr(h^2g^2hg)-\tr(h^2ghg^2)\Big)=\\
&=&\frac{1}{2}\Big(\tr(h^2g^2hg)+\tr(h^2ghg^2)\Big)-\frac{1}{2} T_5(g,h)
\, .
\end{eqnarray*}
By Lemma \ref{hhgghg} the sum $\tr(h^2g^2hg)+\tr(h^2ghg^2)$ can be
expressed as a polynomial in invariants of lower order.
\item $k=7$: There are two types of nontrivial invariants in this case:
\begin{enumerate}
\item $\tr(h^ig^jh^lg^m)$, $i+j+l+m=7$. If one of the powers $i,j,l,m$,
is equal to $3$ or more, we can decrease the order by an appropriate
substitution in equation (\ref{fundrel3}). Next, we observe that
there are only two possible cases when all powers $i,j,l,m$ are smaller
than $3$, namely $\tr(h^2g^2h^2g)$ and $\tr(h^2g^2hg^2)$. Substituting
$h\rightarrow h^2g$ into equation (\ref{fundrel2}) we obtain:
$$
\tr(h^2g^2h^2g)=\tr(g\cdot h^2g\cdot h^2g)=\tr(g\cdot h^2g)\tr(h^2g)
-\overline{\tr(g\cdot h^2g)\tr(g)}+\overline{\tr(g^2\cdot h^2g)}.
$$
Analogously we deal with  $\tr(h^2g^2hg^2)$.

\item $\tr(h^2ghghg)$, $\tr(g^2hghgh)$. By Lemma \ref{indepinv3}
we can express $\tr(h^2ghghg)=\tr(h\cdot(hg)^3)$ in terms
of $\tr(h\cdot(hg)^2)$, $\tr(h\cdot(hg))$, $\tr(h)$ and $\tr(hg)$.
For $\tr(g^2hghgh)\, ,$ we get an analogous expression.
\end{enumerate}
\end{itemize}
Finally, by Lemma \ref{hhgghg}, $T_5(g,h)$ is purely imaginary and
$(T_5(hg))^2$ can be expressed as a polynomial in
$T_1,T_2,T_3,T_4$, so only the sign of $T_5$ remains undetermined.
\qed

%%%%%%%%%%%%%%%%%%%%%%%%%%%%%%%%%%%%%%%%%%%%%%%%%%%%%%%%%%%%%%%%%%%%%%%%%%%%%%%%%%%%%%%%%

\setcounter{equation}{0}
\section{The Configuration Space for $N=1$}
\label{N=1}

%%%%%%%%%%%%%%%%%%%%%%%%%%%%%%%%%%%%%%%%%%%%%%%%%%%%%%%%%%%%%%%%%%%%%%%%%%%%%%%%%%%%%%%%%

Applying the theory outlined above is trivial for $N = 1\, :$
From Lemma \ref{indepinv} we immediately get that the orbit space is
uniquely characterized by the trace function, because it generates the algebra
of invariants.
Here, we will explicitly find the image of the Hilbert mapping
$$
\rho:SU(3)/Ad_{SU(3)} \rightarrow \C\cong\R^2 \, ,
$$
which is simply given by the trace function, $\rho=\tr$.

\begin{figure}[hbt]
  \centering
  \epsfig{file=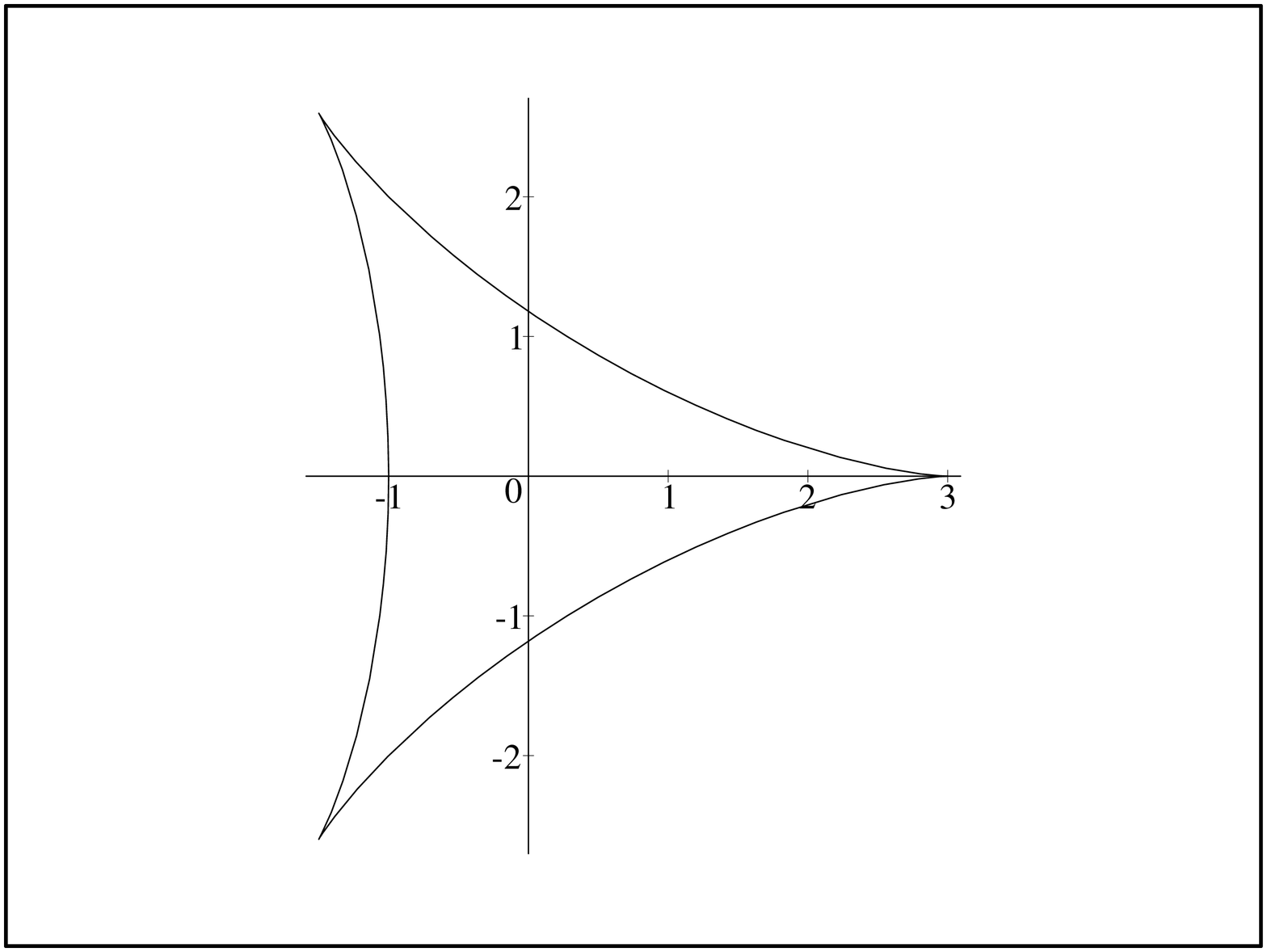,angle=270,width=0.7\textwidth}
  \caption{Hypocycloid.}\label{rys_cykloidy}
\end{figure}

First, observe that the set of possible values of $\tr(g)$, is given
by the sum of the eigenvalues of $g$:
 \begin{equation}
 \label{slad}
\tr(g) \equiv T(\alpha,\beta) = \e^{i\alpha}+\e^{i\beta}+\e^{-i(\alpha+\beta)}, \qquad
\alpha,\beta\in [0,2\pi[.
 \end{equation}
If $g$ belongs to a non-generic orbit of type 2 or 3 in Theorem
\ref{stratN=1},
then at least two eigenvalues are equal.
Thus, setting $\alpha=\beta$ we obtain a curve,
 \begin{equation}
 \label{cycl}
[0,2\pi[ \ni \alpha \mapsto T(\alpha) = 2 \e^{i\alpha} + \e^{-2 i \alpha}
\in \mathbb{C} \, ,
 \end{equation}
which turns out to be a hypocycloid, see Figure \ref{rys_cykloidy}.
We define $\D$ as the compact region enclosed by this curve.
We will show that $\D$ coincides with the image of the Hilbert mapping $\rho$.
For this purpose we first prove the following

\begin{Lemma}\label{dowolna}

Any complex number $T \in \mathbb{C}$ can be presented
in the following form:
 \begin{equation}
 \label{parametryzacja}
T = s\e^{i\theta}+\e^{-2i\theta},
 \end{equation}
where $s\in\mathbb{R}$, $\theta\in[0,\pi[$.

\end{Lemma}

\proof

It is sufficient to show that the mapping
$$
\mathbb{R}\times[0,\pi[ \ni (s,\theta)\mapsto
\phi(s,\theta) :=  s\e^{i\theta}+\e^{-2i\theta}\in\mathbb{C}
$$
is surjective. Denoting $T=t_1+it_2$ we have:
 \begin{equation}
 \label{ukl}
\left\{
\begin{array}{l}
t_1=s\cos\theta+\cos 2\theta \\ \\
t_2=s\sin\theta-\sin 2\theta \\
\end{array}
\right.
 \end{equation}
We show that for given $t_2$, $t_1$ runs over the
whole real axis. For $t_2\neq 0$ ($\sin \theta\neq 0$),
we obtain from the second equation in (\ref{ukl}):
$$
s=\frac{t_2+\sin2\theta}{\sin\theta}.
$$
Substituting this into the first equation of (\ref{ukl}), we get
$t_1$ as a function of $\theta$:
$$
t_1(\theta)=\frac{t_2+\sin 2\theta}{\sin\theta}\cos\theta+\cos
2\theta.
$$
The limits at the boundaries are:
$$
\lim_{\theta\rightarrow0^{+}}t_1(\theta)=\mbox{sgn}(t_2)\cdot\infty \, ,
$$
$$
\lim_{\theta\rightarrow\pi^{-}}t_1(\theta)=-\mbox{sgn}(t_2)\cdot\infty.
$$
The function $\theta\rightarrow t_1(\theta)$ is continuous over
the interval $]0,\pi[$, so it takes all mean values. This means
that for  given $t_2\neq 0$, $t_1\left(\,
]0,\pi[\,\right)=\mathbb{R}$.

For $t_2=0$ we have $\theta=0$. Then, the first of equations
(\ref{ukl}) yields $t_1=s+1$.
\qed

Observe that by substituting
$(\alpha,\beta)\rightarrow(\theta+\phi,\theta-\phi)$ formula
(\ref{slad}) can be rewritten in the form
$$
T(\phi,\theta)=\e^{i(\theta+\phi)}+\e^{i(\theta-\phi)}+\e^{-2i\theta}\,,
$$
yielding
$$
T(\phi,\theta)=(\e^{i\phi}+\e^{-i\phi})\e^{i\theta}+\e^{-2i\theta}
=2\cos\phi \e^{i\theta}+\e^{-2i\theta}
=s\e^{i\theta}+\e^{-2i\theta},
$$
where we have denoted $s:=2\cos\phi$. Thus, in the parametrization
(\ref{parametryzacja}) we have
$$
\D=\left\{T(s,\theta)= s\e^{i\theta}+\e^{-2i\theta}\in{\mathbb C}:
(s,\theta)\in[-2,2]\times[0,\pi[\right\}
$$
and
$$
\partial\D=\left\{T(s,\theta)= s\e^{i\theta}+\e^{-2i\theta}\in{\mathbb C}:
\theta \in[0,\pi[ \ , \ s=2 \ \ {\rm or} \ \ \ s=-2
\right\}.
$$
But
\[
T(-2,\theta) = -2\e^{i\theta}+\e^{-2i\theta} =
2\e^{i(\theta +\pi)}+\e^{-2i(\theta +\pi)} = T(2, \theta + \pi)\ ,
\]
and, whence, $\partial\D$ coincides with the hypocycloid defined above,
$$
\partial\D=\left\{T(\theta)= 2\e^{i\theta}+\e^{-2i\theta}\in{\mathbb C}:
\theta \in[0,2\pi[
\right\}.
$$
One easily checks that in terms of $x=\Re(T)$ and
$y=\Im(T)$,  $\D$
is given by:
 \begin{equation}
 \label{Dxy}
\D =\left\{x+iy\in\mathbb{C}:27-x^4-2x^2y^2-y^4
+8x^3-24xy^2-18x^2-18y^2\geqslant
0\right\} \, .
 \end{equation}

\begin{Theorem}\label{chareqsol}

Let $T \in {\mathbb C}$ and consider the equation
 \begin{equation}
 \label{equ}
\lambda^3- T \lambda^2 + \bar{T}\lambda - 1=0 \,.
 \end{equation}
Its roots $\lambda_1,\lambda_2,\lambda_3$ obey 
 \begin{equation}
 \label{identit--N=1}
 |\lambda_1|=|\lambda_2|=|\lambda_3|=1 \, , \quad
 \lambda_1+\lambda_2+\lambda_3= T \, , \quad
 \lambda_1\lambda_2\lambda_3 = 1 \, ,
 \end{equation}
if and only if $T \in \D \, .$
Consequently, 
$\tr(SU(3))=\D$.

\end{Theorem}

\proof
Using Lemma \ref{dowolna} we can substitute $T(s,\theta) =
s\e^{i\theta}+\e^{-2i\theta}$ into equation (\ref{equ}):
$$
\lambda^3-(s\e^{i\theta}+\e^{-2i\theta})\lambda^2+
(s\e^{-i\theta}+\e^{2i\theta})\lambda-1=0 \, .
$$
It is easy to check that $\lambda_1=\e^{-2i\theta}$ is a root of
this equation. Thus, we can rewrite it in the form:
$$
(\lambda-\e^{-2i\theta})(\lambda^2-s\e^{i\theta}\lambda+\e^{2i\theta})=0.
$$
Let us find the two remaining solutions. For $|s|\leqslant2$
($T\in\D$) we obtain:
 \begin{eqnarray}
 \label{reconstr2}
\lambda_{2,3}&=&\frac{s\pm i\sqrt{4-s^2}}{2}\e^{i\theta},\\ \nonumber
|\lambda_{2,3}|^2&=&\frac{s^2+4-s^2}{4}=1.
 \end{eqnarray}
For $|s|>2$ we get:
 \begin{eqnarray*}
\lambda_{2,3}&=&\frac{s\pm \sqrt{s^2-4}}{2}\e^{i\theta},\\
|\lambda_{2,3}|^2&=&\left(\frac{s^2\pm
\sqrt{s^2-4}}{2}\right)^2\neq 1.
 \end{eqnarray*}
One can check that the sum and the product of roots have the above
properties (in both cases).

Finally, recall that the characteristic polynomial of any
$SU(3)$-matrix is of the form (\ref{charpol}), with eigenvalues
uniquely given as roots of this polynomial. Thus, we have shown
that the numbers $\{\lambda_1,\lambda_2,\lambda_3\}$
are eigenvalues of the characteristic equation of an $SU(3)$-matrix $g$
and (\ref{equ}) coincides with the characteristic equation of $g$
if and only if $\tr(g) \in \D$, so $\tr(SU(3))=\D$.
\koniec

To summarize, combining theorems \ref{stratN=1} and \ref{chareqsol}
we get the following

\begin{Corollary}\label{charN=1}

\noindent
For $N=1\, ,$ the reduced configuration space
$\hat {\cal C}_{\Lambda}$ is isomorphic to $\D $
and contains three orbit types characterized by the
following conditions:

\begin{enumerate}

\item $g$ has three different eigenvalues ~~$\Leftrightarrow$~~ $\tr g$ lies
inside $\D$,

\item $g$ has exactly two different eigenvalues ~~$\Leftrightarrow$~~ $\tr g$
lies on the boundary of $\D$, minus the corners,

\item $g\in \mathcal Z$ ~~$\Leftrightarrow$~~ $\tr g$ is one of the three
corners on the boundary of $\D$.

\end{enumerate}

\end{Corollary}

%%%%%%%%%%%%%%%%%%%%%%%%%%%%%%%%%%%%%%%%%%%%%%%%%%%%%%%%%%%%%%%%%%%%%%%%%%%%%%%%%%%%%%%%%
\setcounter{equation}{0}
\section{The Configuration Space for $N=2$}
\label{N=2}
%%%%%%%%%%%%%%%%%%%%%%%%%%%%%%%%%%%%%%%%%%%%%%%%%%%%%%%%%%%%%%%%%%%%%%%%%%%%%%%%%%%%%%%%%

%%%%%%%%%%%%%%%%%%%%%%%%%%%%%%%%%%%%%%%%%%%%%%%%%%%%%%%%%%%%%%%%%%%%%%%%%%%%%%%%%%%%%

\subsection{Strata in Terms of Invariants}
\label{Invariants and Relations}

%%%%%%%%%%%%%%%%%%%%%%%%%%%%%%%%%%%%%%%%%%%%%%%%%%%%%%%%%%%%%%%%%%%%%%%%%%%%%%%%%%%%%

We define a mapping
$$
\rho=(\rho_1\ldots\rho_9): \mb G^2 \longrightarrow \mathbb{R}^9
$$
by
\begin{eqnarray}
\rho_1(g,h)&:=&\Re(T_1(g,h))=\Re(tr(g)),\\
\rho_2(g,h)&:=&\Im(T_1(g,h))=\Im(tr(g)),\\
\rho_3(g,h)&:=&\Re(T_2(g,h))=\Re(tr(h)),\\
\rho_4(g,h)&:=&\Im(T_2(g,h))=\Im(tr(h)),\\
\rho_5(g,h)&:=&\Re(T_3(g,h))=\Re(tr(hg)),\\
\rho_6(g,h)&:=&\Im(T_3(g,h))=\Im(tr(hg)),\\
\rho_7(g,h)&:=&\Re(T_4(g,h))=\Re(tr(hg)),\\
\rho_8(g,h)&:=&\Im(T_4(g,h))=\Im(tr(hg^2)),\\
\rho_9(g,h)&:=&\Im(T_5(g,h))=\Im(tr(h^2g^2hg)-\tr(h^2ghg^2))\,.
\end{eqnarray}
By Theorem \ref{alginvN=2}, the $\rho_i$ constitute a set of
generators of the algebra of invariant polynomials on $\mb G^2$
with respect to the adjoint action of $G$. According to \cite{schwarz}, the
mapping $\rho$ induces a homeomorphism of $X:=\mb G^2/Ad_G$ onto
the image of $\rho$ in $\mathbb{R}^9\, .$ The set $\{\rho_i\}$ of
generators is, by Theorem \ref{alginvN=2}, subject to a relation,
given in Appendix \ref{app_relation}. We rewrite this relation in
terms of the canonical coordinates $\{x_i \}$ on $\mathbb{R}^9$ by
substituting
$$
tr(g)=x_1+ix_2 \, , \quad tr(h) = x_3+ix_4\, ,\quad tr(hg) =  x_5+ix_6
\, , \quad tr(hg^2)  =  x_7+ix_8
$$
and
$$
\Im(tr(h^2g^2hg)-\tr(h^2ghg^2)) = x_9
$$
into its right-hand-side. By Lemma \ref{hhgghg}, the resulting polynomial
$I_0(\xx)$ is real of order $8$ (it is
of order $4$ in every variable $\xx$). Thus, the relation
defines a hypersurface $Z_1\subset \mathbb{R}$ of codimension $1$
defined by
$$
Z_1:=\left\{(\xxx)\in \mathbb{R}^9: I_0(\xx)=x_9^2\right\}
$$
and the image $\rho(X)$ is a subset of $Z_1 \, .$
On the other hand, by simple dimension counting we know that
$X$ is $8$-dimensional. We conclude that there cannot exist further
independent relations between generators $T_i\, .$ Thus, $\rho(X)$
is an $8$-dimensional compact subset of $Z_1 \, .$
As already mentioned before, in order to identify $\rho(X)$
explicitly, one has to find a number of inequalities between the
above invariants. A full solution of this problem will be presented in a separate paper 
\cite{ChKRS}.

Next, let $X_i$ denote the stratum of $\mb G^2/Ad_G$ corresponding to orbit type $i$.
We are going to characterize each $X_i$ in terms of the above
invariants. We will find a hierarchy of relations: Passing from one stratum to a more
degenerate one, one has to add some new relations to those which
are already fulfilled. This way we obtain a
sequence of algebraic surfaces,
$$
Z_1\supset Z_2\supset Z_3\supset Z_4\supset Z_5 \, ,
$$
characterizing the orbit types. Every $Z_i$ has the property
that the image of $X_i$ under the mapping $\rho$
is a subset of $Z_i$ having the dimension of $Z_i \, .$

According to Theorem \ref{stabilizer2}, a pair $(g,h)$ belongs to
a non-generic stratum, i.e., it has orbit type 2 or higher, iff
$g$ and $h$ have a common eigenvector. The following lemma is due
to I.P.~Volobuev \cite{Igor}:

\begin{Lemma}
\label{LIgor}

The matrices $g$ and $h$ have a common eigenvector if and only if the
following three relations are simultaneously satisfied:
 \begin{eqnarray}\label{nongencond1}
T_5(g,h) & = & 0~,
\\ \label{nongencond2}
\left[g,C+C^{-1}\right] = \left[h,C+C^{-1}\right] & = & 0~,
 \end{eqnarray}
where  $C := hgh^{-1}g^{-1}$ denotes the group commutator. 
\end{Lemma}

\proof
If $x$ is a common eigenvector of $g$ and $h$ then $x$ is an
eigenvector of the commutator $C$ with eigenvalue $1$. Then
the other eigenvalues of $C$ are $\lambda$ and $\overline{\lambda}$, 
for some $\lambda$ obeying $|\lambda|^2 = 1$. In particular, $\tr(C)$ is real. Expressing $\tr(C)$
in terms of generators we obtain
 \begin{eqnarray}\nonumber
\tr(hgh^{-1}g^{-1})
 & = &
\frac{1}{2}\Big(|\tr(g)|^2 + |\tr(h)|^2 + |\tr(hg)|^2 +|\tr(h g^2)|^2
\\ \label{trcom}
 & &
+ \,\,|\tr(g)\tr(hg)|^2 -3 +T_{5}(g,h)\Big)+
\\ \nonumber
 & &
- \,\,  \Re\left(\tr(g)\tr(h)\overline{\tr(hg)}\right)
- \Re\left(\tr(g)\tr(hg)\overline{\tr(hg^2)}\right)~.
\end{eqnarray}
It follows
 \begin{equation}
 \label{trCT5}
\Im(\tr(C)) = \frac{1}{2i}T_5(g,h)~,
 \end{equation}
hence \eqref{nongencond1}. Furthermore, the subspace $E$ orthogonal to $x$ is
an eigenspace of the Hermitean matrix $C + C^{-1}$ with eigenvalue $\lambda+\overline\lambda$. Then
$[g,C+C^{-1}] x = 0$ and $[g,C+C^{-1}] E = 0$, hence \eqref{nongencond2}.
Conversely, assume that \eqref{nongencond1} and \eqref{nongencond2} are
satisfied. According to \eqref{trCT5}, then $\tr(C)$ is real. Due to Lemma
\ref{dowolna}, we can write $\tr(C) = se^{i\theta} + e^{-2i\theta}$. The
rhs.~is real iff $s=2\cos\theta$. Then the reconstruction formula
\eqref{reconstr2} for the eigenvalues of $C$ from $\tr(C)$ implies that $C$ has
an eigenvalue
$$
\lambda_3
 =
\frac{2\cos\theta - i\sqrt{4-4\cos^2\theta}}{2}e^{i\theta}
 =
(\cos\theta - i \sin\theta) e^{i\theta}
 =
1~.
$$
If this eigenvalue is degenerate then $C=\id$, i.e., $g$ and $h$ commute and
therefore have a common eigenvector (even a common eigenbasis). If the
eigenvalue $\lambda_3 = 1$ is nondegenerate then $2$ is a nondegenerate
eigenvalue of $C+C^{-1}$. Let $x$ be a corresponding
eigenvector. According to \eqref{nongencond2},
$$
[g,C+C^{-1}] x = 2gx - (C+C^{-1})gx = 0~,
$$
i.e., $gx$ is again an eigenvector of $C+C^{-1}$ with eigenvalue $2$. It follows
that $x$ is an eigenvector of $g$ and, similarly, of $h$.
 \qed

In terms of invariants, relation \eqref{nongencond2} can be written as
 \begin{eqnarray}
 \label{i1def}
\tr\Big(\left[g,C+C^{-1}\right]\cdot\left[g,C+C^{-1}\right]^\dag\Big)&=&0 \, ,
\\
\label{i2def}
\tr\Big(\left[h,C+C^{-1}\right]\cdot\left[h,C+C^{-1}\right]^\dag\Big)&=&0 \, .
 \end{eqnarray}
We omit the lengthy expressions for these equations in terms of generators.
We only stress that they do not depend on $T_5$. Thus, again using the
canonical coordinate system, we obtain two polynomials $I_1(\xx)$ and $I_2(\xx) \, ,$
which vanish on the nongeneric strata:
$$
Z_2:=\left\{(\xxx)\in Z_1: x_9 = 0, I_1(\xx) = 0, I_2(\xx) = 0\right\}.
$$
The definition of $Z_1$ implies that condition $x_9 = 0$ is
equivalent to $I_0(\xx)=0$, so $Z_2$ can be equivalently viewed as
a subset of $\mathbb{R}^8$ given by equations $I_0=0$, $I_1=0$ and
$I_2=0 \, .$ The image of the generic stratum $X_1$ under the map
$\rho$ then is contained in $Z_1\setminus Z_2$. Hence, inside
$\rho(X)$, it is given by the inequalities
$$
I_0 > 0 \text{~~or~~} I_1 > 0 \text{~~or~~} I_2 > 0~.
$$
One can pass to a set of reduced (with respect to their degree) polynomials
$\{I_0,I_1^R,I_2^R\}\, ,$
\begin{eqnarray}
I_1^R:&=&\frac{1}{2}I_1+I_0 \, ,\\
I_2^R:&=&\frac{1}{2}I_2+I_0 \, ,
\end{eqnarray}
which generate the same ideal in the polynomial algebra, see
Appendix \ref{poly} for their concrete expressions.

The set of orbits of type $3$ or higher consists of pairs of
commuting matrices. The commutativity of a pair $g,h$ can be
expressed in terms of invariants as follows:
$$
\tr(hgh^{-1}g^{-1})-3=0 \, .
$$
Taking the imaginary part yields, according to \eqref{trCT5}, $T_5 = 0 \, .$ 
Denoting 
$$
I_3 = \Re\left(\tr(hgh^{-1}g^{-1})-3\right)\, ,
$$
 we obtain 
$$
I_3 = 0 \, .
$$
$I_3$ can be expressed in terms of $T_1,\ldots,T_4\, ,$
and in terms of canonical coordinates it takes the form
\begin{eqnarray*}
I_3(\xx)&=& {x_{1}}^{2}\,{x_{5}}^{2} + {x_{1}}^{2}\,{x_{6}}^{2}
+{x_{2}}^{2} \,{x_{5}}^{2} + {x_{2}}^{2}\,{x_{6}}^{2}
-2\,{x_{1}}\,{x_{5}}\,{ x_{7}} - 2\,{x_{1}}\,{x_{5}}\,{x_{3}}\\
 & &
-2\,{x_{1}}\,{x_{6}}\,{x_{8}} -2\,{x_{1}}\,{x_{6}}\,{x_{4}}
-2\,{x_{2}}\,{x_{5}}\,{x_{8}} +2\,{x_{2}}\,{x_{5}}\,{x _{4}}
+2\,{x_{2}}\,{x_{6}}\,{x_{7}}\\
 &&
- 2\,{x_{2}}\,{x_{6}}\,{x_{3 }} +{x_{1}}^{2} + {x_{2}}^{2}
+{x_{5}}^{2} + {x_{6}}^{2} +{x_{7}}^{2} + {x_{8}}^{2} +{x_{3}}^{2}
+ {x_{4}}^{2} - 9 \, .
\end{eqnarray*}
Then, the image of the stratum $X_3$ under the mapping $\rho$ is a
subset of
$$
Z_3:=\left\{(\xxx)\in Z_2: I_3(\xx)=0\right\} \, .
$$
Since $\Re\left(\tr(hgh^{-1}g^{-1})-3\right) \leq 0 \, ,$ 
the image of the stratum $X_2$ under $\rho$ is given, as a
subset of $\rho(X)$, by the following equations and inequalities
$$
I_0=0~, ~~~ I_1=0~, ~~~I_2=0~, ~~~I_3 < 0~.
$$

The set of orbits of type $4$ or higher consists of commuting
pairs with a common $2$-dimensional eigenspace. This implies that
both matrices and all their products have degenerate eigenvalues.
The invariants $T_i$, $i=1, \ldots , 4\, ,$ are trace functions of
products of $SU(3)$-matrices, so they take values in $\D\, ,$ see
Theorem \ref{chareqsol}. Thus, by Corollary \ref{charN=1}, the
values of all invariants $T_i$, $i=1, \ldots , 4 \, ,$ computed on
degenerate elements have to belong to $\partial\D$. The polynomial
defining this boundary has the following form, see (\ref{Dxy}):
$$
B(x_1,x_2):=27 - {x_{1}}^{4} - 2\,{x_{1}}^{2}\,{x_{2}}^{2} - {
x_{2}}^{4} + 8\,{x_{1}}^{3} - 24\,{x_{1}}\,{x_{2}}^{2} - 18\,{x_{
1}}^{2} - 18\,{x_{2}}^{2} \, .
$$
Thus, we have
$$
Z_4:=\left\{(\xxx)\in Z_3
 :
B(x_1,x_2)=B(x_3,x_4)=B(x_5,x_6)=B(x_7,x_8)=0\right\}\,.
$$
Accordingly, the image of the stratum $X_3$ under the map $\rho$
is given, as a subset of $\rho(X)$, by the relations
$$
I_0=0~, ~~~ I_1=0~, ~~~I_2=0~, ~~~I_3 = 0
$$
and the inequalities
$$
B(x_1,x_2) > 0 \text{~~~or~~~} B(x_3,x_4) > 0 \text{~~~or~~~} B(x_5,x_6) > 0
\text{~~~or~~~}  B(x_7,x_8) > 0\,.
$$

Finally, the subset of orbits of type $5$ consists of pairs of
matrices belonging to $\mathcal{Z}$. They fulfill
$|\tr(g)|=|\tr(h)|=3$, so we have
$$
Z_5:=\left\{(\xxx)\in Z_4: x_1^2+x_2^2-9=0, \,
x_3^2+x_4^2-9=0\right\}
$$
and the image of the stratum $X_4$ under the map $\rho$ is given,
as a subset of $\rho(X)$, by
$$
I_0=I_1=I_2=I_3=B(x_1,x_2)=B(x_3,x_4)=B(x_5,x_6)=B(x_7,x_8)=0
$$
and
$$
x_1^2 + x_2^2 - 9 < 0 \text{~~~or~~~} x_3^2 + x_4^2 - 9 < 0~.
$$

%%%%%%%%%%%%%%%%%%%%%%%%%%%%%%%%%%%%%%%%%%%%%%%%%%%%%%%%%%%%%%%%%%%%%%%%%%%%%%%%%%%%%%%%%

\subsection{Geometric Structure of Strata}
\label{N=2--geomstr}

%%%%%%%%%%%%%%%%%%%%%%%%%%%%%%%%%%%%%%%%%%%%%%%%%%%%%%%%%%%%%%%%%%%%%%%%%%%%%%%%%%%%%%%%%

In this section we give a description of the strata in terms of subsets and
quotients of $G=SU(3)$ and calculate their dimensions. We use the following
notation. Let $H$ be a subgroup of $G$. Then
 \begin{eqnarray*}
N(H) & := & \text{normalizer of $H$ in $G$,}
\\
\mb G^2_H & := & \text{set of pairs $(g,h)$ with stabilizer $H$,}
\\
\mb G^2_{(H)} & := & \text{set of pairs $(g,h)$ invariant under
$H$,}
\\
\mb G^2_{[H]} & := & \text{set of pairs $(g,h)$ of type $[H]$.}
\end{eqnarray*}
We obviously have $\mb G^2_H \subset \mb G^2_{(H)}$ and 
$\mb G^2_H \subset \mb G^2_{[H]}$. Since we have labelled the orbit types $[H]$ by 
$i=1,\dots,5$, we denote the strata $\mb G^2_{[H]}$ by $\mb G^2_i\, .$  Moreover, in what
follows, the symbol $\setminus$ always means taking the set
theoretical complement, whereas $/$ means taking the quotient.

For orbit type $5$, Theorem \ref{stratyfikacja} immediately yields
that the corresponding stratum is
$$
X_5 = \mc Z \times \mc Z~.
$$
It consists of nine isolated points.

For the remainig orbit types, recall from the basic theory of Lie
group actions \cite{Bredon} that the projection $\pi_i \colon \mb G^2_i\to X_i$ is a 
locally trivial fibre bundle with typical fibre $G/H$ associated with
the $N(H)/H$-principal bundle $\mb G^2_H \to X_i$, which is naturally embedded 
into the associated bundle. Here $H$ is a
representative of the conjugacy class $i$ and we have the following 
diffeomorphism
 \begin{equation}
 \label{Gdiffstrat}
X_i ~~\cong~~ \mb G^2_{H} ~\Big /~ N(H)/H~,
 \end{equation}
where $N(H)/H$ is the right coset group acting by inner automorphisms on 
$\mb G^2_{H}$.
Thus, for each orbit type we have to choose a representative and
then compute the rhs.~ of \eqref{Gdiffstrat}.

We start with orbit type $4$. As a representative, we choose the
subgroup \eqref{stabilizer2}. Let us denote it by $U(2)_1$. We
have
$$
\mb G^2_{U(2)_1}
 ~~=~~
\mb G^2_{(U(2)_1)} ~\Big\backslash~ \mc Z\times\mc Z
$$
and
 \begin{equation}
 \label{typU2}
\mb G^2_{(U(2)_1)} = C(U(2)_1) \times C(U(2)_1) = U(1)_1\times
U(1)_1~,
 \end{equation}
where $C(\cdot)$ denotes the centralizer in $G$ and $U(1)_1$
denotes the subgroup \eqref{Gclfot2}.  Hence,
$$
\mb G^2_{U(2)_1}
 ~~=~~
U(1)_1\times U(1)_1 ~\Big\backslash~ \mc Z\times\mc Z~.
$$
Since $U(2)_1$ and $U(1)_1$ centralize each other, their
normalizers coincide. Since the only way in which $N(U(1)_1)$ can
act on $U(1)_1$ is by a permutation of the entries, it must act
trivially. It follows
$$
N(U(2)_1) = N(U(1)_1) = C(U(1)_1) = U(2)_1~,
$$
and the factorization in \eqref{Gdiffstrat} is trivial. Therefore,
\eqref{Gdiffstrat} yields
$$
X_4
 ~~\cong~~
U(1)_1\times U(1)_1
 ~\Big\backslash~
\mc Z\times\mc Z~.
$$
The dimension of $X_4$ is 2.

As a representative for orbit type $3$ we choose the subgroup
\eqref{stabilizer1} of diagonal matrices. Let us denote it by
$\maxtor$. The set $\mb G^2_{\maxtor}$ consists of the pairs that
are invariant under $\maxtor$ minus those that are of orbit type
$4$ or higher, i.e., that are conjugate to a pair invariant under
$U(2)_1$:
$$
\mb G^2_\maxtor
 ~~= ~~
\mb G^2_{(\maxtor)}
 ~\Big\backslash~
\left(
\bigcup\nolimits_{g\in G} ~g\,\mb G^2_{(U(2)_1)}\,g^{-1}
\right)
~.
$$
We have
 \begin{equation}
 \label{typT}
\mb G^2_{(\maxtor)}
 =
C(\maxtor) \times C(\maxtor)
 =
\maxtor\times\maxtor
 \end{equation}
and, from formula \eqref{typU2},
$$
g\,\mb G^2_{(U(2)_1)}\, g^{-1}
 =
g\,\big(U(1)_1\times U(1)_1\big)\, g^{-1}
 =
\left(g\,U(1)_1\,g^{-1}\right) \times \left(g\,U(1)_1\,g^{-1}\right)~.
$$
Subtraction of this subset from $\maxtor\times \maxtor$ is only
nontrivial if $g U(1)_1 g^{-1} \subseteq \maxtor$. The subgroups
arising this way are $U(1)_1$ as well as
 \begin{eqnarray*}
U(1)_2
 & = &
\{\diag(\beta,\alpha,\beta)
 :
\alpha,\beta\in U(1)\,,\,\beta^2 = \overline\alpha\}~,
\\
U(1)_3
 & = &
\{\diag(\beta,\beta,\alpha)
 :
\alpha,\beta\in U(1)\,,\,\beta^2 = \overline\alpha\}~.
 \end{eqnarray*}
Thus,
$$
\mb G^2_\maxtor
 ~~=~~
\maxtor\times\maxtor
 ~\Big\backslash~
\left(\bigcup\nolimits_{i=1}^3 U(1)_i\times U(1)_i \right)~.
$$
The quotient $N(\maxtor)/\maxtor$ is the Weyl group of $G=SU(3)$,
isomorphic to the permutation group $S_3$. Hence,
$$
X_3
 ~~=~~
 \left(
\maxtor\times\maxtor
 ~\Big\backslash~
\left(\bigcup\nolimits_{i=1}^3 U(1)_i\times U(1)_i \right) 
 \right)
 ~\Big/ ~
S_3~,
$$
where $S_3$ acts on the elements of $\maxtor$ by permuting the
entries. The dimension of the stratum $X_3$ is $4$. Note that if
we take the quotient $(\maxtor\times\maxtor)/S_3$, also the points
of orbit type $4$ and $5$ are factorized in the proper way. One
can make this precise by saying that $(\maxtor\times\maxtor)/S_3$
is isomorphic, as a stratified space, to the subspace
$$
X_3\cup X_4 \cup X_5 \subseteq X = \mb G^2/Ad_G~.
$$
As we will see below, this is not true in general.

Next, consider orbit type $2$. As a representative, we choose the
subgroup $U(1)_1$, given by \eqref{Gclfot2}. Using an argument
analogous to that for orbit type $3$, together with formula
\eqref{typT} and  $C(U(1)_1) = U(2)_1$, we find
$$
\mb G^2_{U(1)_1}
 ~~=~~
U(2)_1 \times U(2)_1
 ~\Big\backslash~
\left(\bigcup\nolimits_{g\in G} g (\maxtor\times\maxtor)
g^{-1}\right)~.
$$
A pair $(g,h)\in U(2)_1\times U(2)_1$ is
conjugate to an element of $\maxtor\times\maxtor$ iff $g$ and $h$
belong to the same maximal toral sugroup in $U(2)_1$. Thus,
$$
\mb G^2_{U(1)_1}
 ~~=~~
U(2)_1 \times U(2)_1
 ~\Big\backslash~
 \left(
 \bigcup\nolimits_{\widetilde\maxtor}
\widetilde\maxtor\times\widetilde\maxtor
 \right)~,
$$
where the union is over all maximal tori in $U(2)_1$. As for the
normalizer, we already know that $N(U(1)_1) = U(2)_1$, hence we
have to factorize by $U(2)_1/U(1)_1 \cong SU(2)$, i.e., by $U(2)_1$ modulo its
center:
$$ 
X_2
 ~~\cong~~
 \left(
U(2)_1 \times U(2)_1
 ~\Big\backslash~
 \left(
 \bigcup\nolimits_{\widetilde\maxtor}
\widetilde\maxtor\times\widetilde\maxtor
 \right)
 \right)
 ~\Big/~
U(2)_1/U(1)_1~.
$$
We see that this stratum has dimension $5$. We remark that in
\eqref{GstratU1} it is important to remove the pairs of higher
symmetry, because they would not be factorized in the proper way
here. Since $U(1)_1$ is the center of $U(2)_1$, we get
\begin{equation}
 \label{GstratU1}
X_2
 ~~\cong~~
 \left(
U(2)_1 \times U(2)_1
 ~\Big\backslash~
 \left(
 \bigcup\nolimits_{\widetilde\maxtor}
\widetilde\maxtor\times\widetilde\maxtor
 \right)
 \right)
 ~\Big/~
U(2)_1~.
 \end{equation}
Moreover, $\bigcup\nolimits_{\widetilde\maxtor} 
\widetilde\maxtor\times\widetilde\maxtor$ contains all non-generic orbit types 
of the $U(2)_1$-action. Hence, the rhs.~of \eqref{GstratU1} is isomorphic to the
generic stratum of the orbit space of the action of the abstract
Lie group $U(2)$ by diagonal conjugation on $U(2)\times U(2)$,
i.e.,
 \begin{equation}
 \label{typU1}
X_2
 ~~\cong~~
\left( (U(2)\times U(2)) ~\Big/~ U(2)\right)_{\rm gen}~.
 \end{equation}
One option to analyze this quotient is to restrict the action to the subgroup
$SU(2)\subset U(2)$ and to rewrite the two factors $U(2)$ using the Lie group
isomorphism
$$
U(2) ~~\cong~~ (U(1)\times SU(2))~\Big/~\Z_2\,,
$$
thus obtaining
$$
(U(2)\times U(2))~\Big/~U(2)
 ~~\cong~~
\Big( U(1)\times U(1)
 \times
\Big(\big(SU(2)\times SU(2)\big) \big/ SU(2)\Big)\Big)
 ~\Big/~
\Big(\Z_2\times\Z_2\Big)~.
$$
Here the quotient $\big(SU(2)\times SU(2)\big) / SU(2)$ is known
as the ''pillow''. It consists of a $3$-dimensional stratum
(corresponding to the interior), a $2$-dimensional stratum (the
boundary minus the $4$ edges) and a $0$-dimensional stratum (the
$4$ edges).

Another option is to apply an algorithm which provides a
decomposition of quotients of diagonal (or joint) actions on
direct product spaces into quotients of the individual factors.
Since we will use this algorithm again to describe the generic
stratum $X_1$ below, we will explain it in some generality. Let
$H$ be a Lie group acting on a manifold $M$ and consider the
diagonal action of $H$ on $M\times M$ (one can easily generalize
the procedure to diagonal action on $M_1 \times\cdots\times M_n$).
In what follows, we denote the sets of orbit types of the action 
of $H$ on $M$, of a subgroup  $K\subseteq H$ on $M$ and of 
$H$ on $M \times M$ by $\mc O(M,H)\, ,$ $\mc O(M,K)$ and 
$\mc O(M \times M,H)\, ,$ respectively. We start with decomposing
$$
(M\times M)~\Big/~H
 ~~=~~
 \bigcup_{[K]\in\mc O(M,H)}
\big(M_{[K]} \times M\big)~\Big/~H~.
$$
If two pairs $(x_1,x_2),(y_1,y_2) \in M_K \times
M\subset M_{[K]}\times M$ are
conjugate under $h\in H$, then conjugation of the stabilizer of
$x_1$ by $h$ yields the stabilizer of $y_1$. Since both are equal
to $K$, $h$ is in the normalizer of $K$ in $H$, $h\in N(K)$. Thus,
$$
\big(M_{[K]} \times M\big)~\Big/~H
 ~~=~~
\big(M_K\times M\big)~\Big/~N(K) \, ,
$$
for some fixed representative $K$ of the orbit type
$[K]$. Factorization by  $N(K)$ can be achieved by first
factorizing by $K$ and then by $N(K)/K$. Since $K$ acts trivially
on the factor $M_K$, we obtain
 \begin{equation}
 \label{GdecoK}
(M\times M)~\Big/~H
 ~~=~~
 \bigcup_{[K]\in\mc O(M,H)}
\big(M_K\times (M/K)\big) ~\Big/~ N(K)/K~.
 \end{equation}
We decompose $M/K$ by orbit types of the $K$-action on $M$:
 \begin{equation}
 \label{GdecoK'}
M/K = \bigcup_{[K']_K\in\mc O(M,K)} \big(M/K\big)_{[K']_K}~.
 \end{equation}
Here $[K']_K$ denotes the conjugacy class of the subgroup
$K'\subseteq K$ in $K$. Inserting \eqref{GdecoK'} into \eqref{GdecoK},
we obtain
 \begin{equation}
 \label{GdecoH}
(M\times M)~\Big/~H
 ~~=
\bigcup_{[K]\in\mc O(M,H)}
 \left(M_K\times
 \left(
\bigcup_{[K']_K\in\mc O(M,K)}
 (M/K)_{[K']_K}
 \right)
 \right)
~\Big/~
 N(K)/K~.
 \end{equation}
Consider, on the other hand, the decomposition of $(M\times M)/H$ by orbit types,
$$
(M\times M)~\Big/~H
 ~~=~~ 
\bigcup_{[L]\in\mc O(M\times M,H)} \left((M\times
M)~\Big/~H\right)_{[L]}~.
$$
A representative of the rhs.\ of \eqref{GdecoH} is given by $(x,y)$, 
where $x\in M_K$ and $y$ can be chosen such that it has orbit type $K'$ under 
the action of $K$. The stabilizer of this pair under the action of $H$ is 
given by intersecting the stabilizer of $x$ under the action of $H$, which is 
$K$, with the stabilizer of $y$ under the action of $H$. The 
intersection yields the stabilizer of $y$ under the action of $K$, which is $K'$. 
Hence, the stabilizer of $(x,y)$ under the action of $H$ is $K'$ and the orbit type is
$[K']$, where the conjugacy class is taken in $H$. Thus, for
every $[L] \in \mc O(M\times M,H)$, we have
 \begin{equation}
 \label{Gdecostrat}
\left((M\times M)~\Big/~H\right)_{[L]}
 ~~=~~
\bigcup_{[K]\in\mc O(M,H)}
 \left(M_K\times
 \left(
\bigcup_{[K']_K\in\mc O(M,K)\atop [K'] = [L]}
 (M/K)_{[K']_K}
 \right)
 \right)
~\Big/~
 N(K)/K~.
 \end{equation}
At this stage, the equality sign just means bijective
correspondence on the level of abstract sets. Of course, this can
be made more precise by saying how the individual manifolds on the
rhs.~are glued together to build up the manifold on the lhs. Here
we do not elaborate on this, for details we refer to \cite{ChKRS}.

Let us apply \eqref{Gdecostrat} to the quotient given by
\eqref{typU1}, i.e. to the case $M=H=U(2)$ with conjugate action.
Representatives of orbit types of the $U(2)$-action on $U(2)$ are
$K=U(2)$ and $K=\maxtor$, where $\maxtor$ denotes the subgroup of
$U(2)$ consisting of diagonal matrices (obviously, if we identify
$U(2)$ with the subgroup $U(2)_1$ of $SU(3)$, this is consistent
with the notation $\maxtor$ used above). Representatives of orbit
types of the $K$-action on $U(2)$ are $K'=U(2)$, $\maxtor$ for $K=U(2)$ and
$K'=\maxtor$, $U(1)$ for $K=\maxtor$.
Here $U(1)$ denotes the center of $U(2)$. Hence, the only piece
in the decomposition \eqref{Gdecostrat} that belongs to the
generic stratum of the $U(2)$-action on $U(2)\times U(2)$ (orbit type
$[U(1)]$) is that labelled by the subgroups $K=\maxtor$ and
$K'=U(1)$. The first factor of this piece is
$$
U(2)_\maxtor
 =
\maxtor\setminus U(1)~,
$$
the second one
$$
\big(U(2)/\maxtor\big)_{[U(1)]_\maxtor}
 =
\big(U(2)/\maxtor\big)_{\rm gen}~.
$$
The quotient group $N(K)/K = N(\maxtor)/\maxtor$ is the Weyl group
of $U(2)$. It is isomorphic to the permutation group $S_2$ and can
be represented on $U(2)$ by conjugation by the permutation matrix
$\left[\begin{array}{cc} 0 & 1 \\ 1 & 0 \end{array}\right]$. Of
course, on the first factor this amounts to interchanging the
entries. Thus, we end up with
$$
X_2
 \cong
\left(\big(U(2)\times U(2)\big) ~\Big/~ U(2) \right)_{\rm gen}
 ~~=~~
 \Big(
\big(\maxtor\setminus U(1)\big)
 \times
\big(U(2)/\maxtor\big)_{\rm gen}
 \Big)
~\Big/~ S_2~.
$$
Clearly, $\big(U(2)/\maxtor\big)_{\rm gen}$ can be further analyzed,
in a similar way as above.

Finally, consider the generic stratum $X_1$. Again, we apply
\eqref{Gdecostrat}, where now $M=H=G=SU(3)$ with conjugate
$SU(3)$-action. Representatives of orbit types of the $G$-action
on $G$ are $K=G$, $U(2)_1$, and $\maxtor$. For $K=G$, the orbit
types of the $K$-action on $G$ are again $[G]$, $[U(2)]$ and
$[\maxtor]$, hence these pieces do not contribute to $X_1$. For
$K=U(2)_1$ and $K=\maxtor$, the $K$-action on $G$ has one orbit
type represented by $\mc Z$. For both actions, this orbit type is
the generic one. Thus, for $X_1$, the decomposition
\eqref{Gdecostrat} consists of one piece labelled by the subgroups
$K=U(2)_1$ and $K' = \mc Z$ and one piece labelled by $K=\maxtor$
and $K' = \mc Z$. Computing these pieces we obtain
$$
X_1
 ~~=~~
\big(U(1)_1\setminus\mc Z\big)
 \times
\big(G/U(2)_1\big)_{\rm gen}
 ~~~\cup~~~
\bigg(\left(\maxtor
 ~\Big\backslash
\left(\bigcup\nolimits_{i=1}^3 ~U(1)_i\right)\right)
 \times
\big(G/\maxtor\big)_{\rm gen}\bigg) ~\Big/~ S_3~,
$$
where the action of the Weyl group $S_3$ on $G$ can be represented by
conjugation by the $3\times 3$-permutation matrices. These are generated,
e.g., by
$$
\left[\begin{array}{ccc}
 1 & 0 & 0 \\ 0 & 0 & 1 \\ 0 & 1 & 0
\end{array}\right]
 ~~,~~
\left[\begin{array}{ccc}
0 & 1 & 0 \\ 1 & 0 & 0 \\ 0 & 0 & 1
\end{array}\right]~.
$$
(Notice that the permutation matrices of negative sign have determinant $-1$,
hence they are not in $SU(3)$.)
On the first factor, $S_3$ acts by permuting the entries.
We note again that the quotients $\big(G/U(2)_1\big)_{\rm gen}$
and $\big(G/\maxtor\big)_{\rm gen}$ can be further analyzed.

%%%%%%%%%%%%%%%%%%%%%%%%%%%%%%%%%%%%%%%%%%%%%%%%%%%%%%%%%%%%%%%%%%%%%%%%%%%%%%%%%%%%%%%%%

\subsection{Representatives of Orbits}
\label{Representatives}

%%%%%%%%%%%%%%%%%%%%%%%%%%%%%%%%%%%%%%%%%%%%%%%%%%%%%%%%%%%%%%%%%%%%%%%%%%%%%%%%%%%%%%%%%

As above, we denote strata by ${\mathbf G}^2_i \subset {\mathbf G}^2\,,$ and the
corresponding pieces of the stratified orbit space by
$X_i = {\mathbf  G}^2_i/Ad_G \subset {\mathbf  G}^2/Ad_G$,
$i = 1, \dots , 5\, .$
In this subsection we present representatives for each orbit type. More precisely,
we define {\em local} cross sections
$$
 X_i \supset {\cal U}_i \ni [{\bf g}] \rightarrow {\bf s}([{\bf g}])
 \equiv (s_1, s_2)([{\bf g}]) \in {\mathbf G}^2_i \, ,
$$
for each bundle
$$
\pi_i  \colon  {\mathbf G}^2_i \to  X_i \, .
$$
Here, ${\cal U}_i$ denotes a dense subset of $X_i \, .$
For that purpose, we use a system of local trivializations of $SU(3)$, viewed as an
$SU(2)$-principal bundle over $S^5$, see Appendix \ref{wiazka}.

%%%%%%%%%%%%%%%%%%%%%%%%%%%%%%%%%%%%%%%%%%%%%%%%%%%%%%%%%%%%%%%%%%%%%%%%%%%%%%%%%%%%%%%%%%

\paragraph{The generic stratum:}

%%%%%%%%%%%%%%%%%%%%%%%%%%%%%%%%%%%%%%%%%%%%%%%%%%%%%%%%%%%%%%%%%%%%%%%%%%%%%%%%%%%%%%%%%%

The projection $\pi_1 \colon  {\bf G}^2_1 \to  X_1$ of the generic
stratum is a locally trivial principal fibre bundle with structure
group $G/\mc Z$. Using arguments developed in \cite{Fleisch} one
can prove that this bundle is non-trivial and that one can find a
system of local trivializations (respectively local cross
sections), defined over a covering of $X_1$ with open subsets,
which are all dense with respect to the natural measure (the one
induced by the Haar-measure).

\begin{Proposition}

There exists a local cross section
$$
 X_1 \supset {\cal U}_1 \ni [{\bf g}] \rightarrow {\bf s}([{\bf g}])
 \equiv (s_1, s_2)([{\bf g}]) \in {\mathbf G}^2_1 \, ,
$$
of the generic stratum with ${\bf s}$ given by
\begin{equation}
\label{genericpair}
 s_1= \left[
 \begin{array}{ccc}
 \lambda_1&0&0\\
 0&\lambda_2&0\\
 0&0&\lambda_3\\
 \end{array}
 \right],
 \quad
 s_2=\macierz{a}{-\delta^{-1}b^{\dag}}{b}{\delta\left(
 \id-\frac{bb^{\dag}}{1+|a|}\right)} \times \left[
\begin{array}{c|rr}
1&0&0\\
\hline
0&c&d\\
0&-\bar{d}&\bar{c}\\
\end{array}\right]
\,,
\end{equation}
where:
\begin{eqnarray}
&&|\lambda_1|=|\lambda_2|=|\lambda_3|=1,
\quad \lambda_1\lambda_2\lambda_3=1,\nonumber\\
&&b=\left[
\begin{array}{c}
b_1\\b_2
\end{array} \right],\quad b_1,b_2\in \mathbb{R}_+,\nonumber\\
&&|a|^2+b_1^2+b_2^2=1,\label{parametry}\\
&&a=|a|\delta^{-2},\nonumber
\\&&|c|^2+|d|^2=1.\nonumber
\end{eqnarray}
\end{Proposition}

\proof
Let
$$
X_1 \supset {\cal U}_1
\ni [{\bf g}] \rightarrow {\bf s}([{\bf g}]) \equiv
(s_1, s_2)([{\bf g}])
\in {\mathbf G}^2_1
$$
be a local cross section, with ${\cal U}_1$ dense in $X_1\, .$
Since $Ad_G$ acts (pointwise) on this cross section, we can fix
the gauge by bringing ${\bf s}$ to a special form. Since $s_1$ and
$s_2$ are in generic position on ${\cal U}_1 \, ,$ they have no
common eigenvector and at least one element of this pair, say
$s_1$, has three different eigenvalues. Thus, on this
neighbourhood, we can fix the gauge in two steps: First, we
diagonalize $s_1$ and next we use the stabilizer of this diagonal
element to bring $s_2$ to a special form. Since $s_1$ and $s_2$
have no common eigenvector, this fixes the (remaining) stabilizer
gauge completely, (up to ${\mathbb Z}_3$). Thus, we can assume
that $s_1$ is diagonal, with eigenvalues ordered in a unique way,
and that $s_2$ has the form, defined by the cross section
\eqref{tryw} in Appendix \ref{wiazka},
\begin{equation}
\label{tryw1}
 s_2 =
  \macierz{a}{-\delta_i b^{\dag}}{b}{\delta_i^{-1} \left(
  {\bf 1}-\frac{bb^{\dag}}{1+|a|}\right)} \times \macierz{1}{0}{0}{S} \,
\, , ~S \in SU(2) \, .
\end{equation}
Let
\[
\pi^{-1}_1({\cal U}_1) \ni (s_1,s_2) \mapsto f(s_1,s_2)  \in G
\]
belong to the stabilizer of $s_1 \, .$ Since $s_1$ is diagonal, $f$
can be written in the form
$$
f=\left[
\begin{array}{ccc}
{\mathrm e}^{-i(\alpha+\beta)}&0&0\\
0&{\mathrm e}^{i\alpha}&0\\
0&0&{\mathrm e}^{i\beta}
\end{array}
\right].
$$
The action of $f$ on an arbitrary group element $g$ is given by:
 \begin{equation}
 \label{adaction}
\left[
\begin{array}{ccc}
g_{11}&g_{12}&g_{13}\\
g_{21}&g_{22}&g_{23}\\
g_{31}&g_{32}&g_{33}\\
\end{array} \right]
\rightarrow \left[
\begin{array}{ccc}
g_{11}&{\mathrm e}^{-i(\alpha+2\beta)}g_{12}&{\mathrm e}^{-i(2\alpha+\beta)}g_{13}\\
{\mathrm e}^{i(\alpha+2\beta)}g_{21}&g_{22}&{\mathrm e}^{-i(\beta-\alpha)}g_{23}\\
{\mathrm e}^{i(2\alpha+\beta)}g_{31}&{\mathrm e}^{i(\beta-\alpha)}g_{32}&g_{33}\\
\end{array} \right].
 \end{equation}
Thus, we can choose the phases $\alpha$ and $\beta$ in such a way that
after transformation with $f$, the entries $b_i$ of $b$ occuring in
\eqref{tryw1} are real and positive.
\qed

By the results of Subsection \ref{Invariants and Relations}, it is clear
that the representative ${\bf s}$ can be expressed in terms of
invariants $t_i := T_i(s_1,s_2)$,
$i=1, \dots,5 \, .$ With some effort, one can find these expressions explicitly.
Here, we only sketch how to do that. In section \ref{N=1} we
have already found the eigenvalues
$\{\lambda_1,\lambda_2,\lambda_3\}$ in terms of $t_1=\tr(s_1)$. Thus, we are left with
calculating $s_2\, .$ For that purpose, denote the diagonal entries of $s_2$ by $x$, $y$
and $z$. Then, we have
\begin{eqnarray}
t_2 & = & x+y+z \, ,\nonumber\\
t_3 & = & \lambda_1 x+\lambda_2 y+\lambda_3 z \, ,\nonumber\\
t_4 & = & \lambda_1^2 x+\lambda_2^2 y+\lambda_3^2 z \, .\nonumber
\end{eqnarray}
This is system of linear equations for $x,y,z$, which can be trivially solved.
The second, non-trivial step consists in expressing the
parameters $a,b,c,d,\delta$ in terms of $x,y,z$,
by solving the set of non-linear equations
\begin{eqnarray}
x & = & a \, ,\nonumber\\
y & = & \delta c-\frac{\delta}{1+|a|}(b_1^2c-b_1b_2\overline{d}) \, ,\label{xyz}\\
z & = & \delta \overline{c}-\frac{\delta}{1+|a|}(b_1b_2d+b_2^2\overline{c}) \, ,\nonumber
\end{eqnarray}
where, of course, relations (\ref{parametry}) have to be taken into
account. It can be shown that this set of equations has two solutions,
corresponding to different parameters $b_1,b_2,d$:
\begin{eqnarray}
a&=&x, \nonumber\\
\delta & = &\sqrt{\frac{|a|}{a}} \, , \nonumber\\
c & = &\frac{\overline{\delta} y+\delta \overline{z}}{1+|a|} \, ,\nonumber\\
{b_1}^{\pm} & = & \frac{1}{\sqrt{2}}\left[
2(c_1q_1+c_2q_2)+(1-|c|^2)(1-|a|^2) \pm \sqrt{\Delta}
\right]^{1/2} \, , \nonumber\\
{b_2}^{\pm} & = & \sqrt{1-|a|^2-b_1^2} \, ,\nonumber \\
{d_1}^{\pm} & = & \frac{c_1b_1^2-q_1}{b_1b_2} \, ,\nonumber \\
{d_2}^{\pm} & = & \frac{-c_2b_1^2+q_2}{b_1b_2} \, , \nonumber
\end{eqnarray}
where
\begin{eqnarray}
c_1&:=&\re(c), \quad c_2:=\im(c)\, ,\nonumber\\
d_1&:=&\re(d), \quad d_2:=\im(d) \, ,\nonumber\\
q_1&:=&-\re(\overline{\delta}y-c)(1+|a|) \, ,\nonumber\\
q_2&:=&-\im(\overline{\delta}y-c)(1+|a|) \, ,\nonumber
\end{eqnarray}
and
$$
\Delta = \left[2(c_1q_1+c_2q_2)+(1-|c|^2)(1-|a|^2)\right]^2-4(q_1^2+q_2^2) \, .
$$
Next, observe that the matrices described by these two sets of parameters are
related, namely one of them is equal to the transposition of
the second one. On the other hand, all invariants
$t_i$, $i=1, \dots, 4\, ,$ are invariant under transposition of
matrices. The two solutions are distinguished by the value of
$T_5(s_1,s_2)$, which has the property
$$
T_5(s_1,s_2)=-T_5(s_1^T,s_2^T) \, .
$$
In terms of matrix elements of $s_1$ and $s_2$, $T_5$ has the following form:
$$
T_5(s_1,s_2)= \pm (\lambda_1-\lambda_2)(\lambda_2-\lambda_3)(\lambda_3-\lambda_1)
\sqrt{\Delta}\, .
$$
Thus, calculating the value of $T_5(s_1,s_2)$ enables us to choose the
correct sign in front of the square root of $\Delta$ and to obtain a unique solution.

%%%%%%%%%%%%%%%%%%%%%%%%%%%%%%%%%%%%%%%%%%%%%%%%%%%%%%%%%%%%%%%%%%%%%%%%%%%%%%%%%%%%%%%%%%%%%%%
\paragraph{The $U(1)$-stratum:}
%%%%%%%%%%%%%%%%%%%%%%%%%%%%%%%%%%%%%%%%%%%%%%%%%%%%%%%%%%%%%%%%%%%%%%%%%%%%%%%%%%%%%%%%%%%%%%%

Let ${\bf s}$ be a local cross section of the (non-trivial) bundle $ \pi_2  \colon {\mathbf
G}^2_2 \to  X_2 \, . $ There exists one common eigenvector of
$s_1$ and $s_2$. Assume that it is the first eigenvector of $s_1
\, .$ After diagonalizing $s_1$, the pair $(s_1,s_2)$ has the
following form
 \begin{equation}
 \label{repr2}
s_1= \left[
\begin{array}{ccc}
\lambda_1&0&0\\
0&\lambda_2&0\\
0&0&\lambda_3\\
\end{array}
\right],\quad s_2= \macierz{\det(S)^{-1}}{0}{0}{S},
 \end{equation}
where $S\in U(2)$. The stabilizer $H_{\bf s} \cong U(1)$ of ${\bf
s}$ is given by \eqref{Gclfot2}. Thus, to obtain a cross section,
we have to fix the $S_2$-action, which permutes the second and
third basis vectors and the $H_{\bf s}$-action on $s_2 \, .$
First, since $\lambda_2 \neq \lambda_3$, these eigenvalues can be
uniquely ordered, for example by increasing phase. Next, the
$H_{\bf s}$-action is fixed by requiring that the left lower entry
of $s_2$ has to be real and positive. Thus, we get the following
local cross section:
 \begin{equation}
 \label{repru1}
s_1= \left[
\begin{array}{ccc}
\lambda_1&0&0\\
0&\lambda_2&0\\
0&0&\lambda_3\\
\end{array}
\right],\quad s_2= \left[
\begin{array}{c|rr}
\delta^{-2}&0&0\\
\hline
0&\delta c&-\delta^2 d\\
0&d&\delta \bar{c}\\
\end{array}\right],
 \end{equation}
where:
 \begin{eqnarray}
&&|\lambda_1|=|\lambda_2|=|\lambda_3|=1,
\quad \lambda_1\lambda_2\lambda_3=1,\nonumber\\
&&|\delta|=1,\nonumber\\
&&|c|^2+d^2=1,\quad d\in \mathbb{R}_+ .\nonumber
 \end{eqnarray}
Again, the representative \eqref{repru1} can be expressed in terms
of invariants: The eigenvalues $\lambda_1,\lambda_2,\lambda_3$ of
$s_1$ are given in terms of $t_1 \, .$ If
$\lambda_1\neq\lambda_2$, we can proceed in the same way as for
the generic stratum above, i.e., by solving the set of equations
(\ref{xyz}). This way, we obtain the diagonal components
$\delta^{-2},\delta c,\delta \bar{c}$ of $s_2$, and we can compute
the coefficients $c$ and $\delta$. There exist two solutions for
$c$ and $\delta$ but they describe the same matrix. If
$\lambda_1=\lambda_2$, equations \eqref{xyz} imply
$$(\delta^{-2}+\delta c)=(x+y) \, , \quad \delta \bar{c}=z \, ,
$$
which can be solved with respect to $c$ and $\delta^2$:
$$
\delta^2=\frac{2}{(x+y)\pm\sqrt{(x+y)^2-4\bar{z}}} \, ,
\quad c=\delta\bar{z} \, .
$$
(There are two values for $\delta^2$, but only one of them satisfies
the condition $|\delta|^2=1$. Taking the square root of the correct one then 
yields two solutions for $\delta$, but these give the same matrix.)
Finally, one calculates
$$
d=\sqrt{1-|c|^2} \, .
$$

%%%%%%%%%%%%%%%%%%%%%%%%%%%%%%%%%%%%%%%%%%%%%%%%%%%%%%%%%%%%%%%%%%%%%%%%%%%%%%%%%%%%%%%%%%%%%%%

\paragraph{The $U(1)\times U(1)$-stratum:}

%%%%%%%%%%%%%%%%%%%%%%%%%%%%%%%%%%%%%%%%%%%%%%%%%%%%%%%%%%%%%%%%%%%%%%%%%%%%%%%%%%%%%%%%%%%%%%%

Let ${\bf s}$ be a local cross section of the (non-trivial) bundle $ \pi_3  \colon
{\mathbf G}^2_3 \to  X_3 \, . $ In this case, $s_1$ and $s_2$ can
be jointly diagonalized:
$$
s_1= \left[
\begin{array}{ccc}
\lambda_1&0&0\\
0&\lambda_2&0\\
0&0&\lambda_3\\
\end{array}
\right],\quad s_2= \left[
\begin{array}{ccc}
\delta_1&0&0\\
0&\delta_2&0\\
0&0&\delta_3\\
\end{array}\right],
$$
where:
\begin{eqnarray*}
&&|\lambda_1|=|\lambda_2|=|\lambda_3|=1,
\quad \lambda_1\lambda_2\lambda_3=1,
\\
&&|\delta_1|=|\delta_2|=|\delta_3|=1, \quad
\delta_1\delta_2\delta_3=1\,.
\end{eqnarray*}
Since there is no common $2$-dimensional
eigenspace, the remainder of the action of the stabilizer $H_{\bf
s} \cong U(1) \times U(1)$ is the permutation group $S_3 \, .$ To
fix the $S_3$-action, observe that, according to Corollary
\ref{pairsandtriples}, either one of the matrices has three
different eigenvalues or both have a pair of degenerate
eigenvalues corresponding to distinct eigenspaces. In the first
case, we can fix the $S_3$-action by ordering the three distinct
eigenvalues. In the second case, we can put the unique
nondegenerate eigenvalue of $s_1$ in the first place and establish
the order of the two remaining eigenvectors by ordering the
corresponding two distinct eigenvalues of $s_2$.

Expressing ${\bf s}$ in terms of invariants is then immediate: All
eigenvalues can be calculated in terms of the traces $t_1 =
\tr(s_1)$ and $t_2 = \tr(s_2)$. To determine which eigenvalues of
$s_1$ and $s_2$ correspond to the same eigenvector it is
sufficient to know the value of $t_3 = \tr(s_1 s_2)$. It can take
six values corresponding to the permutations of the eigenvalues of
$s_2$ relative to those of $s_1$.

%%%%%%%%%%%%%%%%%%%%%%%%%%%%%%%%%%%%%%%%%%%%%%%%%%%%%%%%%%%%%%%%%%%%%%%%%%%%%%%%%%%%%%%%%%%%%%%

\paragraph{The $U(2)$-stratum:}

%%%%%%%%%%%%%%%%%%%%%%%%%%%%%%%%%%%%%%%%%%%%%%%%%%%%%%%%%%%%%%%%%%%%%%%%%%%%%%%%%%%%%%%%%%%%%%%

Let ${\bf s}$ be a cross section of the (trivial) bundle $ \pi_4  \colon {\mathbf
G}^2_4 \to  X_4 \, . $ Obviously, ${\bf s}$ can be taken in the
following form:
$$
s_1= \left[
\begin{array}{ccc}
\lambda_1&0&0\\
0&\lambda_2&0\\
0&0&\lambda_2\\
\end{array}
\right],\quad s_2= \left[
\begin{array}{ccc}
\delta_1&0&0\\
0&\delta_2&0\\
0&0&\delta_2\\
\end{array}\right],
$$
where $|\lambda_1|=|\lambda_2|=|\delta_1|=|\delta_2|=1$,
$\lambda_1\lambda_2^2=\delta_1\delta_2^2=1$. For expressing
$(s_1,s_2)$ in terms of invariants it is sufficient to know the
values $t_1$ and $t_2 \, ,$ because there is only one possible
order.

%%%%%%%%%%%%%%%%%%%%%%%%%%%%%%%%%%%%%%%%%%%%%%%%%%%%%%%%%%%%%%%%%%%%%%%%%%%%%%%%%%%%%%%%%%%%%%%

\paragraph{The $SU(3)$-stratum:}

%%%%%%%%%%%%%%%%%%%%%%%%%%%%%%%%%%%%%%%%%%%%%%%%%%%%%%%%%%%%%%%%%%%%%%%%%%%%%%%%%%%%%%%%%%%%%%%

Let ${\bf s}$ be a cross section of the (trivial) bundle $ \pi_5  \colon
{\mathbf G}^2_5 \to  X_5 \, . $ Then,
$$
s_1= \left[
\begin{array}{ccc}
\lambda&0&0\\
0&\lambda&0\\
0&0&\lambda\\
\end{array}
\right],\quad s_2= \left[
\begin{array}{ccc}
\delta&0&0\\
0&\delta&0\\
0&0&\delta\\
\end{array}\right],
$$
define a unique cross section, with $\lambda^3=1$ and
$\delta^3=1$. The traces of both matrices take one of the
following three values: $3\cdot \e^{i\frac{2k\pi}{3}}$, $k=0,1,2
\, .$ Thus, expressing them in terms of invariants is trivial.

\subsection*{Acknowledgements}

The authors would like to thank I.P.\ Volobuev for discussions on how to derive 
the relations defining the strata. S.\ C.\ is grateful to the Graduiertenkolleg
'Quantenfeldtheorie' at the University of Leipzig for financial support. 
G.\ R.\ and M.\ S.\ acknowledge funding by the German
Research Council (DFG) under the grant RU 692/3-1.

\newpage

%%%%%%%%%%%%%%%%%%%%%%%%%%%%%%%%%%%%%%%%%%%%%%%%%%%%%%%%%%%%%%%%%%%%%%%%%%%%%%%%
\appendix
\renewcommand{\theequation}{\Alph{section}.\arabic{equation}}
%%%%%%%%%%%%%%%%%%%%%%%%%%%%%%%%%%%%%%%%%%%%%%%%%%%%%%%%%%%%%%%%%%%%%%%%%%%%%%%%

%%%%%%%%%%%%%%%%%%%%%%%%%%%%%%%%%%%%%%%%%%%%%%%%%%%%%%%%%%%%%%%%%%%%%%%%%%%
%%%%%%%%%%%%%%%%%%%%%%%%%%%%%%%%%%%%%%%%%%%%%%%%%%%%%%%%%%%%%%%%%%%%%%%%%%%

\setcounter{equation}{0}
\section{A Principal Bundle Atlas for the $SU(3)$ group manifold}
\label{wiazka}

%%%%%%%%%%%%%%%%%%%%%%%%%%%%%%%%%%%%%%%%%%%%%%%%%%%%%%%%%%%%%%%%%%%%%%%%%%%
%%%%%%%%%%%%%%%%%%%%%%%%%%%%%%%%%%%%%%%%%%%%%%%%%%%%%%%%%%%%%%%%%%%%%%%%%%

It is well known that the group $SU(3)$ can be viewed as a
principal bundle over the sphere $S^5$ with structure
group $SU(2)$,
 \begin{equation}
 \label{bundle}
SU(2) \hookrightarrow SU(3) \stackrel{\pi}{\rightarrow}  S^5 \, ,
 \end{equation}
with $\pi$ being the canonical projection from $SU(3)$ onto the right coset space
$SU(3)/SU(2) \cong  S^5$.
An explicit description of $\pi$ is obtained as follows:
Any $3\times 3$ matrix can be written in the form
 \begin{equation}
 \label{postacg}
g=\macierz{a}{c^{\dag}}{b}{B} \, ,
 \end{equation}
with $a\in{\mathbb C},\; b,c\in{\mathbb C}^2$ and a complex $2\times 2$-matrix
$B$. The condition that $g$ belongs to $U(3)$, namely
$$
gg^{\dag}= {\bf 1} = g^{\dag}g,
$$
translates into the following relations for entries of $g$:
\begin{eqnarray}
&|a|^2+\|b\|^2=1=|a|^2+\|c\|^2,&\label{u1}\\
&\bar{a}b+Bc=0=ac+B^{\dag}b,&\label{u2}\\
&bb^{\dag}+BB^{\dag}= {\bf 1} =cc^{\dag}+B^{\dag}B.&\label{u3}
\end{eqnarray}
We embed the subgroup $SU(2)$ of $SU(3)$ as follows:
$$
SU(2) \ni S \rightarrow h = \macierz{1}{0}{0}{S} \in SU(3).
$$
Observe that then $SU(2)$ is the stabilizer of the vector
$$
e_1:=\left[\begin{array}{c}1\\0\\0\end{array}\right] \in S^5
\subset {\mathbb C}^3~.
$$
The image of the left action of $g \in
SU(3)$ on $e_1$ is exactly the first column of $g$, which -- on
the other hand -- is also invariant under right action of $SU(2)$.
Thus, $\pi(g)$ can be identified with the first column of $g$,
$$
\pi(g) = \left[
 \begin{array}{cc}
  a\\
  b_1\\
  b_2
  \end{array}
  \right] \in S^5 \subset {\mathbb C}^3 \ ,
$$
which by (\ref{u1}) has norm $1$, indeed.

Next, we construct an atlas of local trivializations of the bundle
(\ref{bundle}). Observe first that, according to (\ref{u3}),
$\det(B)=0$ iff $\|b\| =1$ and, whence, iff $a=0$. Thus, let us
assume $a \neq 0$ and construct appropriate trivializations of
(\ref{bundle}) over the open set ${\mathcal O} = \{(a,b) | a \ne 0
\} \subset S^5$. Using the polar decomposition $B=AV$, where
$A>0$, $V\in U(2)$, we can rewrite equation (\ref{u3}) as follows:
$$
bb^{\dag}= {\bf 1} - A^2 = V\,c\,c^{\dag}V^{\dag},
$$
yielding
\begin{eqnarray}
c&=& \mathrm{}-\e^{-i\phi}V^{\dag}b,\quad \phi\in{\mathbb R},\label{c}\\
A^2&=& {\bf 1} -bb^{\dag}.\label{Akwadrat}
\end{eqnarray}
Formulae (\ref{u2}) and (\ref{c}) imply
$$
(\bar{a}-\e^{-i\phi}A ) b = 0,
$$
which means that $b$ is an eigenvector of the matrix $A$ with
eigenvalue $\bar{a}\e^{i\phi}$. Positivity of $A$ implies
$|a|=\bar{a}\mathrm{}\e^{i\phi}$.

From equation (\ref{Akwadrat}) we have $A=\sqrt{{\bf 1}
-bb^{\dag}}$. Since $A>0$ this formula defines $A$ uniquely.
Obviously, it must be of the form
 \begin{equation}
 \label{wielomian}
A=\alpha {\bf 1} + \beta \,bb^{\dag}.
 \end{equation}
Plugging this into equation (\ref{Akwadrat}) yields
 \begin{equation}
 \label{A}
A= {\bf 1} -\frac{1}{1+|a|}bb^{\dag}.
 \end{equation}
We conclude that any matrix $g\in U(3)$ which fulfils the condition
$a \neq 0$ can be written in the following form:
 \begin{equation}
 \label{ogolneg}
g=\macierz{a}{-\e^{i\phi}b^{\dag}}{b}{{\bf 1} -\frac{bb^{\dag}}{1+|a|}}
\cdot \macierz{1}{0}{0}{V},
 \end{equation}
with $|a|^2+\|b\|^2=1,\quad a=|a|\e^{i\phi}, \quad V\in U(2)$.

Imposing the condition $\det g=1$ is equivalent to
 \begin{equation}
 \label{wyznacznik}
\det A(a+\e^{i\phi}b^{\dag}A^{-1}b)\det V=1.
 \end{equation}
From (\ref{u1}) and (\ref{A}) we have $\det A=|a|$ and
$A^{-1}b=\frac{1}{|a|}b$. Using this, equation (\ref{wyznacznik})
takes the form:
$$
|a|(a+\e^{+i\phi}\frac{\|b\|^2}{|a|})\det V=1.
$$
Finally, substituting $a=|a|\e^{i\phi}$ and using (\ref{u1}), we
obtain:
$$
\det V = \e^{-i\phi} = \frac{\overline a}{|a|}.
$$

We decompose $V=\delta^{-1} S$, where $S\in SU(2)$ and $\delta^{-2}:=\det
V$, or $\delta^2 = \frac{a}{|a|}$. Of course, $|\delta|=1$. Corresponding
to the two choices of the square root of $\frac{a}{|a|}$, we choose
two open subsets ${\mathcal O}_i \subset {\mathcal O}$,
\begin{eqnarray}
{\mathcal O}_1&:=&
\left\{\left(\begin{array}{c}a\\b_1\\b_2\end{array}\right)\in
{\mathcal O}:
\mathrm{phase}(a)\in]-\pi,\pi[\right\},\nonumber\\
{\mathcal O}_2&:=&
\left\{\left(\begin{array}{c}a\\b_1\\b_2\end{array}\right)\in
{\mathcal O}: \mathrm{phase}(a)\in]0,2\pi[\right\} \, .
\label{otoczenia}
\end{eqnarray}
Then, every element $g \in \pi^{-1}({\mathcal O}_i) \subset SU(3)$,
can be uniquely represented as
$$
g = s_i (\pi (g) ) \cdot h_i(g) \ ,
$$
with $s_i$ being two local cross sections of (\ref{bundle}) over
${\mathcal O}_i$,
 \begin{equation}
 \label{tryw}
  S^5 \supset {\mathcal
  O}_i \ni (a,b) \rightarrow s_i(a,b) =
  \macierz{a}{-\delta_i b^{\dag}}{b}{\delta_i^{-1} \left(
  {\bf 1}-\frac{bb^{\dag}}{1+|a|}\right)} \in SU(3)  \, ,
 \end{equation}
and
 \begin{equation}
 \label{rozklad}
h_i(g) = \macierz{1}{0}{0}{S_i(g)} \subset SU(3)\, \, , \, \, S_i(g) \in SU(2) \ .
 \end{equation}
Thus, corresponding to the two choices of the square root,
we obtain two local bijective mappings
$$
\pi^{-1}({\mathcal O}_i) \ni g \longrightarrow
\chi_i(g) := \left(\pi(g),(s_i (\pi (g) ) )^{-1} \cdot g \right) \in {\mathcal O}_i \times
SU(2)\, .
$$

Similarly, we choose the following open neighborhood of $a=0$:
$$
{\mathcal O}_3:=
\left\{\left(\begin{array}{c}a\\b_1\\b_2\end{array}\right)\in
S^5 : \left(\begin{array}{c}b_1\\b_2\end{array}\right)
\neq\left(\begin{array}{c}0\\0\end{array}\right) \right\} \, .
$$
Then, we find a local cross section $s_3$ over ${\mathcal O}_3$
such that
 \begin{equation}
 \label{u_z_fala}
g = s_3(\pi(g))\cdot h(g) =
\macierz{a}{b^{\dag}}{b}{-{\bf 1} +\frac{1-\bar{a}}{\left\|b\right\|^2}bb^{\dag}}
\cdot \macierz{1}{0}{0}{S(g)},
 \end{equation}
with $S(g) \in SU(2)$, and a local bijective mapping
$$
\pi^{-1}({\mathcal O}_3)\ni g \longrightarrow
\chi_3(g) := \left(\pi(g),(s_3(\pi(g)))^{-1}\cdot g \right)
\in {\mathcal O}_3 \times SU(2)\, .
$$

\begin{Proposition}

The local mappings $\chi_i$, $i=1,2,3$, form an atlas of local
trivializations of the $SU(2)$-principal bundle (\ref{bundle}).

\end{Proposition}

\noindent

\proof

The proof consists of checking the following obvious statements:
\begin{enumerate}
\item
The open neighborhoods ${\mathcal O}_i$ cover $S^5$,
$$
{\mathcal O}_1\cup{\mathcal O}_2\cup{\mathcal O}_3 = S^5  \, .
$$
\item
The mappings
$$
\pi^{-1}({\mathcal O}_i)\ni g \longrightarrow
\chi_i(g) := \left(\pi(g),(s_i(\pi(g)))^{-1}\cdot g \right) \in {\mathcal O}_i \times
SU(2)
$$
are local diffeomorphisms, for $i=1,2,3$.
\item
The mappings $\left\{\chi_i\right\}$ are compatible with the bundle
structure and with the right group action:
\begin{eqnarray}
pr^i_1 \circ \chi_i & = & \pi \, ,\\
(pr^i_2 \circ \chi_i)(g \cdot g') & = & (pr^i_2 \circ \chi_i(g)) \cdot g' \, ,
\end{eqnarray}
for $i=1,2,3$, with $pr^i_{\alpha} \, , \,  \alpha=1,2$, denoting the projection on the
first, respectively second factor of ${\mathcal O}_i \times
SU(2)$.
\end{enumerate}
\qed

\newpage

%%%%%%%%%%%%%%%%%%%%%%%%%%%%%%%%%%%%%%%%%%%%%%%%%%%%%%%%%%%%%%%%%%%%%%%%%%%%%%%%%%%%%%%%%%%%%%%%

\setcounter{equation}{0}
\section{The Relation %of Lemma \ref{hhgghg}/4.}
for $T_5^2$} \label{app_relation}

%%%%%%%%%%%%%%%%%%%%%%%%%%%%%%%%%%%%%%%%%%%%%%%%%%%%%%%%%%%%%%%%%%%%%%%%%%%%%%%%%%%%%%%%%%%%%%%%%

The relation for the square of the invariant $T_5$, referred to in
Lemma \ref{hhgghg}/4., is
\bigskip \\
$\Big(\tr(h^2g^2hg)-\tr(h^2ghg^2)\Big)^2=\\
-27+\tr(h)^2\overline{\tr(h)}^2+18\tr(hg)\overline{\tr(hg)}
+\tr(hg)^2\overline{\tr(hg)}^2+\tr(hg^2)^2\overline{\tr(hg^2)}^2
+18\overline{\tr(hg^2)}\tr(hg^2) \\
-4\tr(h)^3-4\overline{\tr(h)}^3-4\tr(hg^2)^3-4\overline{\tr(g)}^3
-4\overline{\tr(hg^2)}\tr(h)^2\overline{\tr(hg)}
-4\tr(hg^2)\tr(hg)\overline{\tr(h)}^2 \\
-4\tr(hg)^2\tr(hg^2)\overline{\tr(g)}
-4\overline{\tr(hg)}^2\overline{\tr(hg^2)}\tr(g)
-4\tr(hg^2)\overline{\tr(hg)}\tr(g)^2
-6\tr(hg^2)\overline{\tr(hg)}\overline{\tr(g)} \\
-4\overline{\tr(hg)}^2\overline{\tr(hg^2)}\overline{\tr(g)}^2
+8\overline{\tr(hg^2)}^2\overline{\tr(hg)}\overline{\tr(g)}
+\tr(hg^2)^2\overline{\tr(hg)}^2\overline{\tr(g)}^2
+\overline{\tr(hg^2)}^2\tr(hg)^2\tr(g)^2 \\
+8\tr(hg^2)^2\tr(hg)\tr(g)-4\tr(hg)\overline{\tr(hg^2)}\overline{\tr(g)}^2
-4\tr(hg)^2\tr(hg^2)\tr(g)^2-4\tr(hg)^2\overline{\tr(h)}\overline{\tr(hg^2)} \\
+12\overline{\tr(h)}\overline{\tr(hg^2)}\overline{\tr(hg)}
+12\tr(hg^2)\tr(hg)\tr(h)-4\overline{\tr(hg^2)}^2\tr(hg)\tr(h)
-4\tr(hg^2)^2\overline{\tr(h)}\overline{\tr(hg)} \\
-4\overline{\tr(hg^2)}^3
-2\tr(hg^2)\tr(hg)\overline{\tr(hg)}\overline{\tr(hg^2)}
-2\tr(hg^2)\overline{\tr(hg^2)}\tr(h)\overline{\tr(h)}-4\overline{\tr(hg)}^3 \\
 -2\tr(hg^2)\tr(h)\overline{\tr(h)}\overline{\tr(hg)}\overline{\tr(g)}
-4\tr(hg)^3-2\overline{\tr(h)}\overline{\tr(hg)}\tr(h)\tr(hg)
+18\tr(h)\overline{\tr(h)} \\
-2\tr(h)\tr(hg)\overline{\tr(hg)}^2\overline{\tr(g)}\tr(g)^2
-4\overline{\tr(hg)}\tr(h)\overline{\tr(g)}^2
-6\overline{\tr(hg)}\tr(h)\tr(g)
\\
+12\overline{\tr(hg^2)}\tr(h)\overline{\tr(g)}
-4\tr(hg^2)\tr(h)^2\tr(g)-4\tr(hg^2)^2\tr(h)\overline{\tr(g)}
-4\overline{\tr(hg^2)}^2\overline{\tr(h)}\tr(g)
\\
+\overline{\tr(g)}^2\tr(g)^2-2\tr(hg^2)\overline{\tr(hg^2)}\tr(h)
\overline{\tr(hg)}\tr(g)
-2\tr(hg)\tr(hg^2)\overline{\tr(hg)}^2\overline{\tr(g)}^2\tr(g) 
\\
+4\tr(hg^2)\tr(hg)\overline{\tr(hg)}\overline{\tr(hg^2)}
\overline{\tr(g)}\tr(g)
-2\tr(hg)^2\overline{\tr(hg^2)}\overline{\tr(hg)}\overline{\tr(g)}\tr(g)^2 
\\
+2\tr(hg)\tr(hg^2)\overline{\tr(h)}\overline{\tr(hg)}\overline{\tr(g)}^2
+4\tr(hg)\tr(hg^2)\overline{\tr(h)}\overline{\tr(hg)}\tr(g) 
\\
+2\tr(h)\tr(hg)\overline{\tr(hg^2)}\overline{\tr(hg)}\tr(g)^2
+4\tr(h)\tr(hg)\overline{\tr(hg^2)}\overline{\tr(hg)}\overline{\tr(g)} 
\\
+4\tr(hg^2)\tr(hg)\tr(h)\overline{\tr(g)}\tr(g)
+4\overline{\tr(h)}\overline{\tr(hg^2)}\overline{\tr(hg)}
\overline{\tr(g)}\tr(g)
\\
+2\tr(h)\tr(hg^2)\overline{\tr(hg)}^2\overline{\tr(g)}\tr(g) 
+2\tr(hg)^2\overline{\tr(h)}\overline{\tr(hg^2)}\overline{\tr(g)}\tr(g)
+4\tr(hg)^3\overline{\tr(g)}\tr(g)
\\
+4\tr(hg)\overline{\tr(hg)}\overline{\tr(g)}^3
+4\overline{\tr(hg)}^3\overline{\tr(g)}\tr(g) 
-2\overline{\tr(hg^2)}\tr(hg)\tr(h)\overline{\tr(h)}\tr(g)
\\
-6\tr(hg)\overline{\tr(hg^2)}\tr(g)
-4\overline{\tr(hg^2)}\overline{\tr(h)}^2\overline{\tr(g)}
-4\tr(h)\tr(hg)^2\overline{\tr(g)}^2 
-4\overline{\tr(h)}\overline{\tr(hg)}^2\overline{\tr(g)}
\\
-4\tr(hg)\overline{\tr(h)}\tr(g)^2-6\tr(hg)\overline{\tr(h)}\overline{\tr(g)}
-4\tr(h)\tr(hg)^2\tr(g)
+8\overline{\tr(h)}^2\overline{\tr(hg)}\tr(g)
\\
+8\tr(hg)\tr(h)^2\overline{\tr(g)}+\tr(h)^2\overline{\tr(hg)}^2\tr(g)^2
+\tr(hg)^2\overline{\tr(h)}^2\overline{\tr(g)}^2 
-4\tr(hg^2)\overline{\tr(h)}\overline{\tr(g)}^2
\\
 +12\tr(hg^2)\overline{\tr(h)}\tr(g)
 -4\overline{\tr(hg^2)}\tr(h)\tr(g)^2
 -4\overline{\tr(h)}\overline{\tr(hg)}^2\tr(g)^2 
 +4\tr(hg)\overline{\tr(hg)}\tr(g)^3
\\
 -2\tr(hg^2)\overline{\tr(hg^2)}\tr(hg)\overline{\tr(h)}\overline{\tr(g)}
 -2\tr(hg)^2\overline{\tr(h)}\overline{\tr(hg)}\overline{\tr(g)}^2\tr(g) 
\\
 +4\overline{\tr(h)}\overline{\tr(hg)}\tr(h)\tr(hg)\overline{\tr(g)}\tr(g)
 -2\tr(h)\overline{\tr(h)}\overline{\tr(g)}\tr(g)
 -4\tr(g)^3
\\
 -2\overline{\tr(hg^2)}\tr(hg^2)\overline{\tr(g)}\tr(g) 
 -2\tr(hg^2)\overline{\tr(hg^2)}^2\tr(hg)\tr(g)
 -2\tr(hg^2)^2\overline{\tr(hg^2)}\overline{\tr(hg)}\overline{\tr(g)}
\\
 +2\tr(hg^2)\overline{\tr(hg)}\overline{\tr(g)}^2\tr(g) 
 +2\tr(hg)\overline{\tr(hg^2)}\overline{\tr(g)}\tr(g)^2
 +2\tr(hg)\tr(hg^2)\overline{\tr(hg)}^2\overline{\tr(g)}
\\
 +2\tr(hg)^2\overline{\tr(hg^2)}\overline{\tr(hg)}\tr(g) 
 +\tr(hg)^2\overline{\tr(hg)}^2\overline{\tr(g)}^2\tr(g)^2
 -2\tr(hg)^2\overline{\tr(hg)}^2\overline{\tr(g)}\tr(g)
\\
 -2\tr(hg)\overline{\tr(hg)}\overline{\tr(g)}^2\tr(g)^2 
 -8\tr(hg)\overline{\tr(hg)}\overline{\tr(g)}\tr(g)
 +2\tr(h)\tr(hg)\overline{\tr(hg)}^2\tr(g)
\\
 +2\tr(hg)\overline{\tr(h)}\overline{\tr(g)}^2\tr(g) 
 +2\overline{\tr(hg)}\tr(h)\overline{\tr(g)}\tr(g)^2
 +2\tr(hg)^2\overline{\tr(h)}\overline{\tr(hg)}\overline{\tr(g)}
\\
 -2\tr(h)^2\overline{\tr(h)}\overline{\tr(hg)}\tr(g) 
 -2\tr(hg)\tr(h)\overline{\tr(h)}^2\overline{\tr(g)}
 +18\overline{\tr(g)}\tr(g)
 -4\tr(h)\tr(hg^2)\overline{\tr(hg)}^2
$.

\bigskip

\noindent It can be derived in the following way. Consider the
invariant functions $\tr(hghgghghhggh)$ and $\tr(hghgghhgghgh)$ of
order $12$. The sum of them can be expressed in terms of
generators $T_1,\ldots,T_5$ in two different ways. First, we use
the trace identity (\ref{fti}) for $k=4$ and $g_1=gh$, $g_2=gg$,
$g_3=hg$, $g_4=hhgghh$ to express $\tr(hghgghghhggh)$ in terms of
traces of lower order. Next, we use the trace identity (\ref{fti})
for $k=4$ and $g_1=hh$, $g_2=gh$, $g_3=gghhgg$, $g_4=hg$ to
express $\tr(hghgghhgghgh)$. It turns out that in both cases
(which are actually equivalent, because one is obtained from the
other by interchanging $g$ with $h$), we obtain expressions which
can be simplified using standard techniques from Section
\ref{invariantsorbspace}. The final expressions in terms of
generators do not depend on $T_5$.

On the other hand we
observe that the sum
$$
\tr(hghgghghhggh)+\tr(hghgghhgghgh)=
\tr\!\left( (hg)^2\!(gh)^2\! (hg) (gh)\right)+
\tr\!\left((hg)^2\! (gh) (hg) (gh)^2\right)
$$
can be expressed in terms of invariants of lower order using
formula (\ref{symtrid}) (we substitute $h\rightarrow hg$,
$g\rightarrow gh$). In this case, we obtain
a different formula containing $T_5^2$. Taking the difference of
these two expressions yields the above relation.

All calculations described above were made by a computer program
written under \emph{Maple 8.00}. It is worth mentioning that this
program automatically generates polynomial expression in terms of
generators for any trace function (at least up to order $12$)
using only standard techniques, namely fundamental trace
identities and appropriate substitutions in the Cayley equation.

Finally, let us mention that, once the relation has been found, it can be checked by
direct calculation.

\newpage

%%%%%%%%%%%%%%%%%%%%%%%%%%%%%%%%%%%%%%%%%%%%%%%%%%%%%%%%%%%%%%%%%%%%%%%%

\section{The Polynomials $I_0 \, ,$ $I^R_1$ and $I^R_2$}
\label{poly}

%%%%%%%%%%%%%%%%%%%%%%%%%%%%%%%%%%%%%%%%%%%%%%%%%%%%%%%%%%%%%%%%%%%%%%%%

\begin{small}

$I_0(\xxfull)=
\\
\mbox{} - {x_{1}}^{4}\,{x_{5}}^{4} - 2\,{x_{1}}^{4}\,{x_{5}}^{2}\,{x_{6}
}^{2} - {x_{1}}^{4}\,{x_{6}}^{4} - 2\,{x_{1}}^{2}\,{x_{2}}^{2}\,{
x_{5}}^{4} - 4\,{x_{1}}^{2}\,{x_{2}}^{2}\,{x_{5}}^{2}\,{x_{6}}^{2
} - 2\,{x_{1}}^{2}\,{x_{2}}^{2}\,{x_{6}}^{4} - {x_{2}}^{4}\,{x_{5
}}^{4} 
\\
\mbox{} - 2\,{x_{2}}^{4}\,{x_{5}}^{2}\,{x_{6}}^{2} - {x_{2}}^{4}
\,{x_{6}}^{4} + 4\,{x_{1}}^{3}\,{x_{3}}\,{x_{5}}^{3} + 4\,{x_{1}}
^{3}\,{x_{3}}\,{x_{5}}\,{x_{6}}^{2} + 4\,{x_{1}}^{3}\,{x_{4}}\,{x
_{5}}^{2}\,{x_{6}} + 4\,{x_{1}}^{3}\,{x_{4}}\,{x_{6}}^{3} \\
\mbox{} + 4\,{x_{1}}^{3}\,{x_{5}}^{3}\,{x_{7}} + 4\,{x_{1}}^{3}\,
{x_{5}}^{2}\,{x_{6}}\,{x_{8}} + 4\,{x_{1}}^{3}\,{x_{5}}\,{x_{6}}
^{2}\,{x_{7}} + 4\,{x_{1}}^{3}\,{x_{6}}^{3}\,{x_{8}} + 4\,{x_{1}}
^{2}\,{x_{2}}\,{x_{3}}\,{x_{5}}^{2}\,{x_{6}} \\
\mbox{} + 4\,{x_{1}}^{2}\,{x_{2}}\,{x_{3}}\,{x_{6}}^{3} - 4\,{x_{
1}}^{2}\,{x_{2}}\,{x_{4}}\,{x_{5}}^{3} - 4\,{x_{1}}^{2}\,{x_{2}}
\,{x_{4}}\,{x_{5}}\,{x_{6}}^{2} + 4\,{x_{1}}^{2}\,{x_{2}}\,{x_{5}
}^{3}\,{x_{8}} \\
\mbox{} - 4\,{x_{1}}^{2}\,{x_{2}}\,{x_{5}}^{2}\,{x_{6}}\,{x_{7}}
 + 4\,{x_{1}}^{2}\,{x_{2}}\,{x_{5}}\,{x_{6}}^{2}\,{x_{8}} - 4\,{x
_{1}}^{2}\,{x_{2}}\,{x_{6}}^{3}\,{x_{7}} + 4\,{x_{1}}\,{x_{2}}^{2
}\,{x_{3}}\,{x_{5}}^{3} \\
\mbox{} + 4\,{x_{1}}\,{x_{2}}^{2}\,{x_{3}}\,{x_{5}}\,{x_{6}}^{2}
 + 4\,{x_{1}}\,{x_{2}}^{2}\,{x_{4}}\,{x_{5}}^{2}\,{x_{6}} + 4\,{x
_{1}}\,{x_{2}}^{2}\,{x_{4}}\,{x_{6}}^{3} + 4\,{x_{1}}\,{x_{2}}^{2
}\,{x_{5}}^{3}\,{x_{7}} \\
\mbox{} + 4\,{x_{1}}\,{x_{2}}^{2}\,{x_{5}}^{2}\,{x_{6}}\,{x_{8}}
 + 4\,{x_{1}}\,{x_{2}}^{2}\,{x_{5}}\,{x_{6}}^{2}\,{x_{7}} + 4\,{x
_{1}}\,{x_{2}}^{2}\,{x_{6}}^{3}\,{x_{8}} + 4\,{x_{2}}^{3}\,{x_{3}
}\,{x_{5}}^{2}\,{x_{6}} + 4\,{x_{2}}^{3}\,{x_{3}}\,{x_{6}}^{3}
 \\
\mbox{} - 4\,{x_{2}}^{3}\,{x_{4}}\,{x_{5}}^{3} - 4\,{x_{2}}^{3}\,
{x_{4}}\,{x_{5}}\,{x_{6}}^{2} + 4\,{x_{2}}^{3}\,{x_{5}}^{3}\,{x_{
8}} - 4\,{x_{2}}^{3}\,{x_{5}}^{2}\,{x_{6}}\,{x_{7}} + 4\,{x_{2}}
^{3}\,{x_{5}}\,{x_{6}}^{2}\,{x_{8}} \\
\mbox{} - 4\,{x_{2}}^{3}\,{x_{6}}^{3}\,{x_{7}} + 2\,{x_{1}}^{4}\,
{x_{5}}^{2} + 2\,{x_{1}}^{4}\,{x_{6}}^{2} + 4\,{x_{1}}^{2}\,{x_{2
}}^{2}\,{x_{5}}^{2} + 4\,{x_{1}}^{2}\,{x_{2}}^{2}\,{x_{6}}^{2} -
6\,{x_{1}}^{2}\,{x_{3}}^{2}\,{x_{5}}^{2} \\
\mbox{} - 2\,{x_{1}}^{2}\,{x_{3}}^{2}\,{x_{6}}^{2} - 8\,{x_{1}}^{
2}\,{x_{3}}\,{x_{4}}\,{x_{5}}\,{x_{6}} - 8\,{x_{1}}^{2}\,{x_{3}}
\,{x_{5}}^{2}\,{x_{7}} - 8\,{x_{1}}^{2}\,{x_{3}}\,{x_{5}}\,{x_{6}
}\,{x_{8}} - 2\,{x_{1}}^{2}\,{x_{4}}^{2}\,{x_{5}}^{2} \\
\mbox{} - 6\,{x_{1}}^{2}\,{x_{4}}^{2}\,{x_{6}}^{2} - 8\,{x_{1}}^{
2}\,{x_{4}}\,{x_{5}}\,{x_{6}}\,{x_{7}} - 8\,{x_{1}}^{2}\,{x_{4}}
\,{x_{6}}^{2}\,{x_{8}} + 2\,{x_{1}}^{2}\,{x_{5}}^{4} + 4\,{x_{1}}
^{2}\,{x_{5}}^{2}\,{x_{6}}^{2} \\
\mbox{} - 6\,{x_{1}}^{2}\,{x_{5}}^{2}\,{x_{7}}^{2} - 2\,{x_{1}}^{
2}\,{x_{5}}^{2}\,{x_{8}}^{2} - 8\,{x_{1}}^{2}\,{x_{5}}\,{x_{6}}\,
{x_{7}}\,{x_{8}} + 2\,{x_{1}}^{2}\,{x_{6}}^{4} - 2\,{x_{1}}^{2}\,
{x_{6}}^{2}\,{x_{7}}^{2} \\
\mbox{} - 6\,{x_{1}}^{2}\,{x_{6}}^{2}\,{x_{8}}^{2} - 8\,{x_{1}}\,
{x_{2}}\,{x_{3}}^{2}\,{x_{5}}\,{x_{6}} + 8\,{x_{1}}\,{x_{2}}\,{x
_{3}}\,{x_{4}}\,{x_{5}}^{2} - 8\,{x_{1}}\,{x_{2}}\,{x_{3}}\,{x_{4
}}\,{x_{6}}^{2} \\
\mbox{} - 8\,{x_{1}}\,{x_{2}}\,{x_{3}}\,{x_{5}}^{2}\,{x_{8}} - 8
\,{x_{1}}\,{x_{2}}\,{x_{3}}\,{x_{6}}^{2}\,{x_{8}} + 8\,{x_{1}}\,{
x_{2}}\,{x_{4}}^{2}\,{x_{5}}\,{x_{6}} + 8\,{x_{1}}\,{x_{2}}\,{x_{
4}}\,{x_{5}}^{2}\,{x_{7}} \\
\mbox{} + 8\,{x_{1}}\,{x_{2}}\,{x_{4}}\,{x_{6}}^{2}\,{x_{7}} - 8
\,{x_{1}}\,{x_{2}}\,{x_{5}}^{2}\,{x_{7}}\,{x_{8}} + 8\,{x_{1}}\,{
x_{2}}\,{x_{5}}\,{x_{6}}\,{x_{7}}^{2} - 8\,{x_{1}}\,{x_{2}}\,{x_{
5}}\,{x_{6}}\,{x_{8}}^{2} \\
\mbox{} + 8\,{x_{1}}\,{x_{2}}\,{x_{6}}^{2}\,{x_{7}}\,{x_{8}} + 2
\,{x_{2}}^{4}\,{x_{5}}^{2} + 2\,{x_{2}}^{4}\,{x_{6}}^{2} - 2\,{x
_{2}}^{2}\,{x_{3}}^{2}\,{x_{5}}^{2} - 6\,{x_{2}}^{2}\,{x_{3}}^{2}
\,{x_{6}}^{2} \\
\mbox{} + 8\,{x_{2}}^{2}\,{x_{3}}\,{x_{4}}\,{x_{5}}\,{x_{6}} - 8
\,{x_{2}}^{2}\,{x_{3}}\,{x_{5}}\,{x_{6}}\,{x_{8}} + 8\,{x_{2}}^{2
}\,{x_{3}}\,{x_{6}}^{2}\,{x_{7}} - 6\,{x_{2}}^{2}\,{x_{4}}^{2}\,{
x_{5}}^{2} - 2\,{x_{2}}^{2}\,{x_{4}}^{2}\,{x_{6}}^{2} \\
\mbox{} + 8\,{x_{2}}^{2}\,{x_{4}}\,{x_{5}}^{2}\,{x_{8}} - 8\,{x_{
2}}^{2}\,{x_{4}}\,{x_{5}}\,{x_{6}}\,{x_{7}} + 2\,{x_{2}}^{2}\,{x
_{5}}^{4} + 4\,{x_{2}}^{2}\,{x_{5}}^{2}\,{x_{6}}^{2} - 2\,{x_{2}}
^{2}\,{x_{5}}^{2}\,{x_{7}}^{2} \\
\mbox{} - 6\,{x_{2}}^{2}\,{x_{5}}^{2}\,{x_{8}}^{2} + 8\,{x_{2}}^{
2}\,{x_{5}}\,{x_{6}}\,{x_{7}}\,{x_{8}} + 2\,{x_{2}}^{2}\,{x_{6}}
^{4} - 6\,{x_{2}}^{2}\,{x_{6}}^{2}\,{x_{7}}^{2} - 2\,{x_{2}}^{2}
\,{x_{6}}^{2}\,{x_{8}}^{2} \\
\mbox{} - 4\,{x_{1}}^{3}\,{x_{3}}\,{x_{5}} - 4\,{x_{1}}^{3}\,{x_{
4}}\,{x_{6}} - 8\,{x_{1}}^{3}\,{x_{5}}^{2} - 4\,{x_{1}}^{3}\,{x_{
5}}\,{x_{7}} - 8\,{x_{1}}^{3}\,{x_{6}}^{2} - 4\,{x_{1}}^{3}\,{x_{
6}}\,{x_{8}} \\
\mbox{} - 4\,{x_{1}}^{2}\,{x_{2}}\,{x_{3}}\,{x_{6}} + 4\,{x_{1}}
^{2}\,{x_{2}}\,{x_{4}}\,{x_{5}} - 4\,{x_{1}}^{2}\,{x_{2}}\,{x_{5}
}\,{x_{8}} + 4\,{x_{1}}^{2}\,{x_{2}}\,{x_{6}}\,{x_{7}} + 8\,{x_{1
}}^{2}\,{x_{3}}\,{x_{5}}^{2} \\
\mbox{} - 8\,{x_{1}}^{2}\,{x_{3}}\,{x_{5}}\,{x_{7}} - 8\,{x_{1}}
^{2}\,{x_{3}}\,{x_{6}}^{2} + 8\,{x_{1}}^{2}\,{x_{3}}\,{x_{6}}\,{x
_{8}} - 16\,{x_{1}}^{2}\,{x_{4}}\,{x_{5}}\,{x_{6}} + 8\,{x_{1}}^{
2}\,{x_{4}}\,{x_{5}}\,{x_{8}} \\
\mbox{} + 8\,{x_{1}}^{2}\,{x_{4}}\,{x_{6}}\,{x_{7}} - 8\,{x_{1}}
^{2}\,{x_{5}}^{3} + 8\,{x_{1}}^{2}\,{x_{5}}^{2}\,{x_{7}} + 24\,{x
_{1}}^{2}\,{x_{5}}\,{x_{6}}^{2} - 16\,{x_{1}}^{2}\,{x_{5}}\,{x_{6
}}\,{x_{8}} \\
\mbox{} - 8\,{x_{1}}^{2}\,{x_{6}}^{2}\,{x_{7}} - 4\,{x_{1}}\,{x_{
2}}^{2}\,{x_{3}}\,{x_{5}} - 4\,{x_{1}}\,{x_{2}}^{2}\,{x_{4}}\,{x
_{6}} + 24\,{x_{1}}\,{x_{2}}^{2}\,{x_{5}}^{2} - 4\,{x_{1}}\,{x_{2
}}^{2}\,{x_{5}}\,{x_{7}} \\
\mbox{} + 24\,{x_{1}}\,{x_{2}}^{2}\,{x_{6}}^{2} - 4\,{x_{1}}\,{x
_{2}}^{2}\,{x_{6}}\,{x_{8}} + 32\,{x_{1}}\,{x_{2}}\,{x_{3}}\,{x_{
5}}\,{x_{6}} + 16\,{x_{1}}\,{x_{2}}\,{x_{4}}\,{x_{5}}^{2} - 16\,{
x_{1}}\,{x_{2}}\,{x_{4}}\,{x_{6}}^{2} \\
\mbox{} - 16\,{x_{1}}\,{x_{2}}\,{x_{5}}^{2}\,{x_{8}} - 32\,{x_{1}
}\,{x_{2}}\,{x_{5}}\,{x_{6}}\,{x_{7}} + 16\,{x_{1}}\,{x_{2}}\,{x
_{6}}^{2}\,{x_{8}} + 4\,{x_{1}}\,{x_{3}}^{3}\,{x_{5}} + 4\,{x_{1}
}\,{x_{3}}^{2}\,{x_{4}}\,{x_{6}} \\
\mbox{} + 4\,{x_{1}}\,{x_{3}}^{2}\,{x_{5}}\,{x_{7}} + 4\,{x_{1}}
\,{x_{3}}^{2}\,{x_{6}}\,{x_{8}} + 4\,{x_{1}}\,{x_{3}}\,{x_{4}}^{2
}\,{x_{5}} - 4\,{x_{1}}\,{x_{3}}\,{x_{5}}^{3} - 8\,{x_{1}}\,{x_{3
}}\,{x_{5}}^{2}\,{x_{7}} \\
\mbox{} - 4\,{x_{1}}\,{x_{3}}\,{x_{5}}\,{x_{6}}^{2} + 4\,{x_{1}}
\,{x_{3}}\,{x_{5}}\,{x_{7}}^{2} + 4\,{x_{1}}\,{x_{3}}\,{x_{5}}\,{
x_{8}}^{2} - 8\,{x_{1}}\,{x_{3}}\,{x_{6}}^{2}\,{x_{7}} + 4\,{x_{1
}}\,{x_{4}}^{3}\,{x_{6}} \\
\mbox{} + 4\,{x_{1}}\,{x_{4}}^{2}\,{x_{5}}\,{x_{7}} + 4\,{x_{1}}
\,{x_{4}}^{2}\,{x_{6}}\,{x_{8}} - 4\,{x_{1}}\,{x_{4}}\,{x_{5}}^{2
}\,{x_{6}} - 8\,{x_{1}}\,{x_{4}}\,{x_{5}}^{2}\,{x_{8}} - 4\,{x_{1
}}\,{x_{4}}\,{x_{6}}^{3} \\
\mbox{} - 8\,{x_{1}}\,{x_{4}}\,{x_{6}}^{2}\,{x_{8}} + 4\,{x_{1}}
\,{x_{4}}\,{x_{6}}\,{x_{7}}^{2} + 4\,{x_{1}}\,{x_{4}}\,{x_{6}}\,{
x_{8}}^{2} - 4\,{x_{1}}\,{x_{5}}^{3}\,{x_{7}} - 4\,{x_{1}}\,{x_{5
}}^{2}\,{x_{6}}\,{x_{8}} \\
\mbox{} - 4\,{x_{1}}\,{x_{5}}\,{x_{6}}^{2}\,{x_{7}} + 4\,{x_{1}}
\,{x_{5}}\,{x_{7}}^{3} + 4\,{x_{1}}\,{x_{5}}\,{x_{7}}\,{x_{8}}^{2
} - 4\,{x_{1}}\,{x_{6}}^{3}\,{x_{8}} + 4\,{x_{1}}\,{x_{6}}\,{x_{7
}}^{2}\,{x_{8}} + 4\,{x_{1}}\,{x_{6}}\,{x_{8}}^{3} \\
\mbox{} - 4\,{x_{2}}^{3}\,{x_{3}}\,{x_{6}} + 4\,{x_{2}}^{3}\,{x_{
4}}\,{x_{5}} - 4\,{x_{2}}^{3}\,{x_{5}}\,{x_{8}} + 4\,{x_{2}}^{3}
\,{x_{6}}\,{x_{7}} - 8\,{x_{2}}^{2}\,{x_{3}}\,{x_{5}}^{2} - 8\,{x
_{2}}^{2}\,{x_{3}}\,{x_{5}}\,{x_{7}} \\
\mbox{} + 8\,{x_{2}}^{2}\,{x_{3}}\,{x_{6}}^{2} + 8\,{x_{2}}^{2}\,
{x_{3}}\,{x_{6}}\,{x_{8}} + 16\,{x_{2}}^{2}\,{x_{4}}\,{x_{5}}\,{x
_{6}} + 8\,{x_{2}}^{2}\,{x_{4}}\,{x_{5}}\,{x_{8}} + 8\,{x_{2}}^{2
}\,{x_{4}}\,{x_{6}}\,{x_{7}} \\
\mbox{} - 8\,{x_{2}}^{2}\,{x_{5}}^{3} - 8\,{x_{2}}^{2}\,{x_{5}}^{
2}\,{x_{7}} + 24\,{x_{2}}^{2}\,{x_{5}}\,{x_{6}}^{2} + 16\,{x_{2}}
^{2}\,{x_{5}}\,{x_{6}}\,{x_{8}} + 8\,{x_{2}}^{2}\,{x_{6}}^{2}\,{x
_{7}} + 4\,{x_{2}}\,{x_{3}}^{3}\,{x_{6}} \\
\mbox{} - 4\,{x_{2}}\,{x_{3}}^{2}\,{x_{4}}\,{x_{5}} + 4\,{x_{2}}
\,{x_{3}}^{2}\,{x_{5}}\,{x_{8}} - 4\,{x_{2}}\,{x_{3}}^{2}\,{x_{6}
}\,{x_{7}} + 4\,{x_{2}}\,{x_{3}}\,{x_{4}}^{2}\,{x_{6}} - 4\,{x_{2
}}\,{x_{3}}\,{x_{5}}^{2}\,{x_{6}} \\
\mbox{} + 8\,{x_{2}}\,{x_{3}}\,{x_{5}}^{2}\,{x_{8}} - 4\,{x_{2}}
\,{x_{3}}\,{x_{6}}^{3} + 8\,{x_{2}}\,{x_{3}}\,{x_{6}}^{2}\,{x_{8}
} + 4\,{x_{2}}\,{x_{3}}\,{x_{6}}\,{x_{7}}^{2} + 4\,{x_{2}}\,{x_{3
}}\,{x_{6}}\,{x_{8}}^{2} \\
\mbox{} - 4\,{x_{2}}\,{x_{4}}^{3}\,{x_{5}} + 4\,{x_{2}}\,{x_{4}}
^{2}\,{x_{5}}\,{x_{8}} - 4\,{x_{2}}\,{x_{4}}^{2}\,{x_{6}}\,{x_{7}
} + 4\,{x_{2}}\,{x_{4}}\,{x_{5}}^{3} - 8\,{x_{2}}\,{x_{4}}\,{x_{5
}}^{2}\,{x_{7}} \\
\mbox{} + 4\,{x_{2}}\,{x_{4}}\,{x_{5}}\,{x_{6}}^{2} - 4\,{x_{2}}
\,{x_{4}}\,{x_{5}}\,{x_{7}}^{2} - 4\,{x_{2}}\,{x_{4}}\,{x_{5}}\,{
x_{8}}^{2} - 8\,{x_{2}}\,{x_{4}}\,{x_{6}}^{2}\,{x_{7}} - 4\,{x_{2
}}\,{x_{5}}^{3}\,{x_{8}} \\
\mbox{} + 4\,{x_{2}}\,{x_{5}}^{2}\,{x_{6}}\,{x_{7}} - 4\,{x_{2}}
\,{x_{5}}\,{x_{6}}^{2}\,{x_{8}} + 4\,{x_{2}}\,{x_{5}}\,{x_{7}}^{2
}\,{x_{8}} + 4\,{x_{2}}\,{x_{5}}\,{x_{8}}^{3} + 4\,{x_{2}}\,{x_{6
}}^{3}\,{x_{7}} - 4\,{x_{2}}\,{x_{6}}\,{x_{7}}^{3} \\
\mbox{} - 4\,{x_{2}}\,{x_{6}}\,{x_{7}}\,{x_{8}}^{2} - {x_{1}}^{4}
 - 2\,{x_{1}}^{2}\,{x_{2}}^{2} + 2\,{x_{1}}^{2}\,{x_{3}}^{2} + 8
\,{x_{1}}^{2}\,{x_{3}}\,{x_{5}} + 8\,{x_{1}}^{2}\,{x_{3}}\,{x_{7}
} + 2\,{x_{1}}^{2}\,{x_{4}}^{2} \\
\mbox{} + 8\,{x_{1}}^{2}\,{x_{4}}\,{x_{6}} + 8\,{x_{1}}^{2}\,{x_{
4}}\,{x_{8}} + 8\,{x_{1}}^{2}\,{x_{5}}^{2} + 8\,{x_{1}}^{2}\,{x_{
5}}\,{x_{7}} + 8\,{x_{1}}^{2}\,{x_{6}}^{2} + 8\,{x_{1}}^{2}\,{x_{
6}}\,{x_{8}} + 2\,{x_{1}}^{2}\,{x_{7}}^{2} \\
\mbox{} + 2\,{x_{1}}^{2}\,{x_{8}}^{2} - 16\,{x_{1}}\,{x_{2}}\,{x
_{3}}\,{x_{6}} + 16\,{x_{1}}\,{x_{2}}\,{x_{3}}\,{x_{8}} + 16\,{x
_{1}}\,{x_{2}}\,{x_{4}}\,{x_{5}} - 16\,{x_{1}}\,{x_{2}}\,{x_{4}}
\,{x_{7}} \\
\mbox{} - 16\,{x_{1}}\,{x_{2}}\,{x_{5}}\,{x_{8}} + 16\,{x_{1}}\,{
x_{2}}\,{x_{6}}\,{x_{7}} - 16\,{x_{1}}\,{x_{3}}^{2}\,{x_{5}} + 8
\,{x_{1}}\,{x_{3}}^{2}\,{x_{7}} + 32\,{x_{1}}\,{x_{3}}\,{x_{4}}\,
{x_{6}} \\
\mbox{} - 16\,{x_{1}}\,{x_{3}}\,{x_{4}}\,{x_{8}} + 8\,{x_{1}}\,{x
_{3}}\,{x_{5}}^{2} - 8\,{x_{1}}\,{x_{3}}\,{x_{6}}^{2} + 8\,{x_{1}
}\,{x_{3}}\,{x_{7}}^{2} - 8\,{x_{1}}\,{x_{3}}\,{x_{8}}^{2} + 16\,
{x_{1}}\,{x_{4}}^{2}\,{x_{5}} \\
\mbox{} - 8\,{x_{1}}\,{x_{4}}^{2}\,{x_{7}} - 16\,{x_{1}}\,{x_{4}}
\,{x_{5}}\,{x_{6}} - 16\,{x_{1}}\,{x_{4}}\,{x_{7}}\,{x_{8}} + 8\,
{x_{1}}\,{x_{5}}^{2}\,{x_{7}} - 16\,{x_{1}}\,{x_{5}}\,{x_{6}}\,{x
_{8}} \\
\mbox{} - 16\,{x_{1}}\,{x_{5}}\,{x_{7}}^{2} + 16\,{x_{1}}\,{x_{5}
}\,{x_{8}}^{2} - 8\,{x_{1}}\,{x_{6}}^{2}\,{x_{7}} + 32\,{x_{1}}\,
{x_{6}}\,{x_{7}}\,{x_{8}} - {x_{2}}^{4} + 2\,{x_{2}}^{2}\,{x_{3}}
^{2} \\
\mbox{} - 8\,{x_{2}}^{2}\,{x_{3}}\,{x_{5}} - 8\,{x_{2}}^{2}\,{x_{
3}}\,{x_{7}} + 2\,{x_{2}}^{2}\,{x_{4}}^{2} - 8\,{x_{2}}^{2}\,{x_{
4}}\,{x_{6}} - 8\,{x_{2}}^{2}\,{x_{4}}\,{x_{8}} + 8\,{x_{2}}^{2}
\,{x_{5}}^{2} \\
\mbox{} - 8\,{x_{2}}^{2}\,{x_{5}}\,{x_{7}} + 8\,{x_{2}}^{2}\,{x_{
6}}^{2} - 8\,{x_{2}}^{2}\,{x_{6}}\,{x_{8}} + 2\,{x_{2}}^{2}\,{x_{
7}}^{2} + 2\,{x_{2}}^{2}\,{x_{8}}^{2} - 16\,{x_{2}}\,{x_{3}}^{2}
\,{x_{6}} \\
\mbox{} - 8\,{x_{2}}\,{x_{3}}^{2}\,{x_{8}} - 32\,{x_{2}}\,{x_{3}}
\,{x_{4}}\,{x_{5}} - 16\,{x_{2}}\,{x_{3}}\,{x_{4}}\,{x_{7}} - 16
\,{x_{2}}\,{x_{3}}\,{x_{5}}\,{x_{6}} + 16\,{x_{2}}\,{x_{3}}\,{x_{
7}}\,{x_{8}} \\
\mbox{} + 16\,{x_{2}}\,{x_{4}}^{2}\,{x_{6}} + 8\,{x_{2}}\,{x_{4}}
^{2}\,{x_{8}} - 8\,{x_{2}}\,{x_{4}}\,{x_{5}}^{2} + 8\,{x_{2}}\,{x
_{4}}\,{x_{6}}^{2} + 8\,{x_{2}}\,{x_{4}}\,{x_{7}}^{2} - 8\,{x_{2}
}\,{x_{4}}\,{x_{8}}^{2} \\
\mbox{} + 8\,{x_{2}}\,{x_{5}}^{2}\,{x_{8}} + 16\,{x_{2}}\,{x_{5}}
\,{x_{6}}\,{x_{7}} + 32\,{x_{2}}\,{x_{5}}\,{x_{7}}\,{x_{8}} - 8\,
{x_{2}}\,{x_{6}}^{2}\,{x_{8}} + 16\,{x_{2}}\,{x_{6}}\,{x_{7}}^{2}
 \\
\mbox{} - 16\,{x_{2}}\,{x_{6}}\,{x_{8}}^{2} - {x_{3}}^{4} - 2\,{x
_{3}}^{2}\,{x_{4}}^{2} + 2\,{x_{3}}^{2}\,{x_{5}}^{2} + 8\,{x_{3}}
^{2}\,{x_{5}}\,{x_{7}} + 2\,{x_{3}}^{2}\,{x_{6}}^{2} - 8\,{x_{3}}
^{2}\,{x_{6}}\,{x_{8}} \\
\mbox{} + 2\,{x_{3}}^{2}\,{x_{7}}^{2} + 2\,{x_{3}}^{2}\,{x_{8}}^{
2} + 16\,{x_{3}}\,{x_{4}}\,{x_{5}}\,{x_{8}} + 16\,{x_{3}}\,{x_{4}
}\,{x_{6}}\,{x_{7}} + 8\,{x_{3}}\,{x_{5}}^{2}\,{x_{7}} + 16\,{x_{
3}}\,{x_{5}}\,{x_{6}}\,{x_{8}} \\
\mbox{} + 8\,{x_{3}}\,{x_{5}}\,{x_{7}}^{2} - 8\,{x_{3}}\,{x_{5}}
\,{x_{8}}^{2} - 8\,{x_{3}}\,{x_{6}}^{2}\,{x_{7}} + 16\,{x_{3}}\,{
x_{6}}\,{x_{7}}\,{x_{8}} - {x_{4}}^{4} + 2\,{x_{4}}^{2}\,{x_{5}}
^{2} - 8\,{x_{4}}^{2}\,{x_{5}}\,{x_{7}} \\
\mbox{} + 2\,{x_{4}}^{2}\,{x_{6}}^{2} + 8\,{x_{4}}^{2}\,{x_{6}}\,
{x_{8}} + 2\,{x_{4}}^{2}\,{x_{7}}^{2} + 2\,{x_{4}}^{2}\,{x_{8}}^{
2} - 8\,{x_{4}}\,{x_{5}}^{2}\,{x_{8}} + 16\,{x_{4}}\,{x_{5}}\,{x
_{6}}\,{x_{7}} \\
\mbox{} + 16\,{x_{4}}\,{x_{5}}\,{x_{7}}\,{x_{8}} + 8\,{x_{4}}\,{x
_{6}}^{2}\,{x_{8}} - 8\,{x_{4}}\,{x_{6}}\,{x_{7}}^{2} + 8\,{x_{4}
}\,{x_{6}}\,{x_{8}}^{2} - {x_{5}}^{4} - 2\,{x_{5}}^{2}\,{x_{6}}^{
2} + 2\,{x_{5}}^{2}\,{x_{7}}^{2} \\
\mbox{} + 2\,{x_{5}}^{2}\,{x_{8}}^{2} - {x_{6}}^{4} + 2\,{x_{6}}
^{2}\,{x_{7}}^{2} + 2\,{x_{6}}^{2}\,{x_{8}}^{2} - {x_{7}}^{4} - 2
\,{x_{7}}^{2}\,{x_{8}}^{2} - {x_{8}}^{4} + 8\,{x_{1}}^{3} - 24\,{
x_{1}}\,{x_{2}}^{2} \\
\mbox{} + 12\,{x_{1}}\,{x_{3}}\,{x_{5}} - 24\,{x_{1}}\,{x_{3}}\,{
x_{7}} + 12\,{x_{1}}\,{x_{4}}\,{x_{6}} - 24\,{x_{1}}\,{x_{4}}\,{x
_{8}} + 12\,{x_{1}}\,{x_{5}}\,{x_{7}} + 12\,{x_{1}}\,{x_{6}}\,{x
_{8}} \\
\mbox{} + 12\,{x_{2}}\,{x_{3}}\,{x_{6}} + 24\,{x_{2}}\,{x_{3}}\,{
x_{8}} - 12\,{x_{2}}\,{x_{4}}\,{x_{5}} - 24\,{x_{2}}\,{x_{4}}\,{x
_{7}} + 12\,{x_{2}}\,{x_{5}}\,{x_{8}} - 12\,{x_{2}}\,{x_{6}}\,{x
_{7}} \\
\mbox{} + 8\,{x_{3}}^{3} - 24\,{x_{3}}\,{x_{4}}^{2} - 24\,{x_{3}}
\,{x_{5}}\,{x_{7}} + 24\,{x_{3}}\,{x_{6}}\,{x_{8}} + 24\,{x_{4}}
\,{x_{5}}\,{x_{8}} + 24\,{x_{4}}\,{x_{6}}\,{x_{7}} + 8\,{x_{5}}^{
3} \\
\mbox{} - 24\,{x_{5}}\,{x_{6}}^{2} + 8\,{x_{7}}^{3} - 24\,{x_{7}}
\,{x_{8}}^{2} - 18\,{x_{1}}^{2} - 18\,{x_{2}}^{2} - 18\,{x_{3}}^{
2} - 18\,{x_{4}}^{2} - 18\,{x_{5}}^{2} - 18\,{x_{6}}^{2} \\
\mbox{} - 18\,{x_{7}}^{2} - 18\,{x_{8}}^{2} + 27 $

\bigskip

\noindent
$I_1^R(\xxfull)=\\
\mbox{} ~~~\;27 - 9\,{x_{8}}^{2} - 9\,{x_{7}}^{2} - 9\,{x_{3}}^{2} - 9\,{x_{4}
}^{2} - 6\,{x_{2}}\,{x_{6}}\,{x_{7}} + 2\,{x_{7}}^{3} - 6\,{x_{7}
}\,{x_{8}}^{2} - 9\,{x_{6}}^{2} - 9\,{x_{5}}^{2} + 6\,{x_{1}}\,{x
_{5}}\,{x_{7}} \\
\mbox{} + 6\,{x_{2}}\,{x_{5}}\,{x_{8}} - 8\,{x_{1}}\,{x_{2}}\,{x
_{3}}\,{x_{6}} + 8\,{x_{1}}\,{x_{2}}\,{x_{4}}\,{x_{5}} + 2\,{x_{5
}}^{3} - 6\,{x_{5}}\,{x_{6}}^{2} + 4\,{x_{1}}^{2}\,{x_{3}}\,{x_{5
}} - 4\,{x_{2}}^{2}\,{x_{3}}\,{x_{5}} \\
\mbox{} + 6\,{x_{1}}\,{x_{3}}\,{x_{5}} + 6\,{x_{2}}\,{x_{3}}\,{x
_{6}} - 6\,{x_{2}}\,{x_{4}}\,{x_{5}} + 4\,{x_{1}}^{2}\,{x_{4}}\,{
x_{6}} - 4\,{x_{2}}^{2}\,{x_{4}}\,{x_{6}} + 2\,{x_{3}}^{3} - 6\,{
x_{3}}\,{x_{4}}^{2} \\
\mbox{} + 4\,{x_{1}}\,{x_{2}}\,{x_{4}}\,{x_{5}}^{2} - 4\,{x_{2}}
\,{x_{3}}\,{x_{5}}\,{x_{6}} - 4\,{x_{1}}^{2}\,{x_{4}}\,{x_{5}}\,{
x_{6}} + 4\,{x_{2}}^{2}\,{x_{4}}\,{x_{5}}\,{x_{6}} - 4\,{x_{1}}\,
{x_{4}}\,{x_{5}}\,{x_{6}} \\
\mbox{} - 4\,{x_{1}}\,{x_{2}}\,{x_{4}}\,{x_{6}}^{2} - 4\,{x_{1}}
^{2}\,{x_{5}}\,{x_{6}}\,{x_{8}} + 4\,{x_{2}}^{2}\,{x_{5}}\,{x_{6}
}\,{x_{8}} + 4\,{x_{1}}\,{x_{4}}^{2}\,{x_{5}} + 2\,{x_{1}}^{2}\,{
x_{3}}\,{x_{6}}\,{x_{8}} \\
\mbox{} + 2\,{x_{2}}^{2}\,{x_{3}}\,{x_{6}}\,{x_{8}} + 2\,{x_{1}}
^{2}\,{x_{4}}\,{x_{6}}\,{x_{7}} + 2\,{x_{2}}^{2}\,{x_{4}}\,{x_{6}
}\,{x_{7}} + 2\,{x_{1}}^{2}\,{x_{4}}\,{x_{5}}\,{x_{8}} + 2\,{x_{2
}}^{2}\,{x_{4}}\,{x_{5}}\,{x_{8}} \\
\mbox{} + 4\,{x_{1}}^{2}\,{x_{5}}\,{x_{7}} + 6\,{x_{1}}\,{x_{4}}
\,{x_{6}} - 4\,{x_{2}}^{2}\,{x_{5}}\,{x_{7}} + 4\,{x_{1}}^{2}\,{x
_{6}}\,{x_{8}} - 4\,{x_{2}}^{2}\,{x_{6}}\,{x_{8}} - 4\,{x_{1}}\,{
x_{3}}^{2}\,{x_{5}} \\
\mbox{} - 4\,{x_{2}}\,{x_{3}}^{2}\,{x_{6}} + 4\,{x_{2}}\,{x_{4}}
^{2}\,{x_{6}} + 8\,{x_{1}}\,{x_{2}}\,{x_{6}}\,{x_{7}} - 8\,{x_{1}
}\,{x_{2}}\,{x_{5}}\,{x_{8}} - 2\,{x_{1}}^{2}\,{x_{3}}\,{x_{5}}\,
{x_{7}} \\
\mbox{} - 2\,{x_{2}}^{2}\,{x_{3}}\,{x_{5}}\,{x_{7}} + {x_{1}}^{4}
\,{x_{6}}^{2} + {x_{2}}^{4}\,{x_{6}}^{2} - 4\,{x_{1}}^{3}\,{x_{6}
}^{2} + 2\,{x_{1}}^{2}\,{x_{2}}^{2}\,{x_{5}}^{2} + {x_{2}}^{4}\,{
x_{5}}^{2} - 4\,{x_{1}}^{3}\,{x_{5}}^{2} \\
\mbox{} + 2\,{x_{1}}\,{x_{3}}^{2}\,{x_{7}} - 2\,{x_{1}}\,{x_{4}}
^{2}\,{x_{7}} - 2\,{x_{2}}\,{x_{3}}^{2}\,{x_{8}} + 2\,{x_{2}}\,{x
_{4}}^{2}\,{x_{8}} + 2\,{x_{1}}\,{x_{3}}\,{x_{7}}^{2} - 2\,{x_{1}
}\,{x_{3}}\,{x_{8}}^{2} \\
\mbox{} + 2\,{x_{2}}\,{x_{4}}\,{x_{7}}^{2} - 2\,{x_{2}}\,{x_{4}}
\,{x_{8}}^{2} + {x_{1}}^{2}\,{x_{7}}^{2} + {x_{2}}^{2}\,{x_{7}}^{
2} + {x_{1}}^{2}\,{x_{8}}^{2} + {x_{2}}^{2}\,{x_{8}}^{2} - 8\,{x
_{1}}\,{x_{2}}\,{x_{4}}\,{x_{7}} \\
\mbox{} + 8\,{x_{1}}\,{x_{2}}\,{x_{3}}\,{x_{8}} - 4\,{x_{2}}\,{x
_{3}}\,{x_{4}}\,{x_{7}} - 4\,{x_{1}}\,{x_{3}}\,{x_{4}}\,{x_{8}}
 - 4\,{x_{1}}\,{x_{4}}\,{x_{7}}\,{x_{8}} + 4\,{x_{2}}\,{x_{3}}\,{
x_{7}}\,{x_{8}} \\
\mbox{} - 8\,{x_{1}}\,{x_{2}}\,{x_{5}}\,{x_{6}}\,{x_{7}} + 4\,{x
_{1}}^{2}\,{x_{3}}\,{x_{7}} - 4\,{x_{2}}^{2}\,{x_{3}}\,{x_{7}} -
12\,{x_{1}}\,{x_{3}}\,{x_{7}} - 12\,{x_{2}}\,{x_{4}}\,{x_{7}} +
12\,{x_{2}}\,{x_{3}}\,{x_{8}} \\
\mbox{} + 4\,{x_{1}}^{2}\,{x_{4}}\,{x_{8}} - 4\,{x_{2}}^{2}\,{x_{
4}}\,{x_{8}} - 12\,{x_{1}}\,{x_{4}}\,{x_{8}} + 12\,{x_{1}}\,{x_{2
}}^{2}\,{x_{5}}^{2} + 2\,{x_{1}}^{2}\,{x_{2}}^{2}\,{x_{6}}^{2} +
12\,{x_{1}}\,{x_{2}}^{2}\,{x_{6}}^{2} \\
\mbox{} - 2\,{x_{1}}^{3}\,{x_{5}}\,{x_{7}} - 2\,{x_{1}}\,{x_{2}}
^{2}\,{x_{5}}\,{x_{7}} - 2\,{x_{1}}^{2}\,{x_{2}}\,{x_{5}}\,{x_{8}
} + 2\,{x_{1}}^{2}\,{x_{2}}\,{x_{6}}\,{x_{7}} - 2\,{x_{1}}\,{x_{2
}}^{2}\,{x_{6}}\,{x_{8}} \\
\mbox{} + 8\,{x_{1}}\,{x_{6}}\,{x_{7}}\,{x_{8}} + 8\,{x_{2}}\,{x
_{5}}\,{x_{7}}\,{x_{8}} + 4\,{x_{2}}\,{x_{5}}\,{x_{6}}\,{x_{7}}
 - 4\,{x_{1}}\,{x_{2}}\,{x_{5}}^{2}\,{x_{8}} + 4\,{x_{1}}\,{x_{2}
}\,{x_{6}}^{2}\,{x_{8}} \\
\mbox{} - 4\,{x_{1}}\,{x_{5}}\,{x_{6}}\,{x_{8}} - 4\,{x_{1}}\,{x
_{5}}\,{x_{7}}^{2} + 4\,{x_{1}}\,{x_{5}}\,{x_{8}}^{2} + 4\,{x_{2}
}\,{x_{6}}\,{x_{7}}^{2} - 4\,{x_{2}}\,{x_{6}}\,{x_{8}}^{2} - 2\,{
x_{2}}^{3}\,{x_{5}}\,{x_{8}} \\
\mbox{} + 2\,{x_{2}}^{3}\,{x_{6}}\,{x_{7}} - 2\,{x_{1}}^{3}\,{x_{
6}}\,{x_{8}} + 2\,{x_{1}}\,{x_{5}}^{2}\,{x_{7}} - 2\,{x_{1}}\,{x
_{6}}^{2}\,{x_{7}} + 2\,{x_{2}}\,{x_{5}}^{2}\,{x_{8}} - 2\,{x_{2}
}\,{x_{6}}^{2}\,{x_{8}} \\
\mbox{} + 4\,{x_{1}}^{2}\,{x_{5}}^{2} + 4\,{x_{2}}^{2}\,{x_{5}}^{
2} + 4\,{x_{1}}^{2}\,{x_{6}}^{2} + 4\,{x_{2}}^{2}\,{x_{6}}^{2} +
2\,{x_{2}}\,{x_{4}}\,{x_{6}}^{2} + 2\,{x_{1}}^{2}\,{x_{5}}^{2}\,{
x_{7}} - 2\,{x_{2}}^{2}\,{x_{5}}^{2}\,{x_{7}} \\
\mbox{} - 2\,{x_{1}}^{2}\,{x_{6}}^{2}\,{x_{7}} + 2\,{x_{2}}^{2}\,
{x_{6}}^{2}\,{x_{7}} + 8\,{x_{1}}\,{x_{2}}\,{x_{3}}\,{x_{5}}\,{x
_{6}} + 2\,{x_{1}}^{2}\,{x_{3}}\,{x_{5}}^{2} - 2\,{x_{2}}^{2}\,{x
_{3}}\,{x_{5}}^{2} + 2\,{x_{1}}\,{x_{3}}\,{x_{5}}^{2} \\
\mbox{} - 2\,{x_{2}}\,{x_{4}}\,{x_{5}}^{2} - 2\,{x_{1}}^{2}\,{x_{
3}}\,{x_{6}}^{2} + 2\,{x_{2}}^{2}\,{x_{3}}\,{x_{6}}^{2} - 2\,{x_{
1}}\,{x_{3}}\,{x_{6}}^{2} + 8\,{x_{1}}\,{x_{3}}\,{x_{4}}\,{x_{6}}
 - 8\,{x_{2}}\,{x_{3}}\,{x_{4}}\,{x_{5}} \\
\mbox{} + 6\,{x_{3}}\,{x_{6}}\,{x_{8}} + 6\,{x_{4}}\,{x_{5}}\,{x
_{8}} - 6\,{x_{3}}\,{x_{5}}\,{x_{7}} + 6\,{x_{4}}\,{x_{6}}\,{x_{7
}} - 2\,{x_{1}}^{3}\,{x_{4}}\,{x_{6}} - 2\,{x_{1}}\,{x_{2}}^{2}\,
{x_{3}}\,{x_{5}} \\
\mbox{} - 2\,{x_{1}}^{2}\,{x_{2}}\,{x_{3}}\,{x_{6}} + 2\,{x_{1}}
^{2}\,{x_{2}}\,{x_{4}}\,{x_{5}} - 2\,{x_{1}}\,{x_{2}}^{2}\,{x_{4}
}\,{x_{6}} + {x_{1}}^{2}\,{x_{3}}^{2} + {x_{2}}^{2}\,{x_{3}}^{2}
 - 2\,{x_{1}}^{3}\,{x_{3}}\,{x_{5}} \\
\mbox{} - 2\,{x_{2}}^{3}\,{x_{3}}\,{x_{6}} + 2\,{x_{2}}^{3}\,{x_{
4}}\,{x_{5}} + 6\,{x_{1}}^{2}\,{x_{5}}\,{x_{6}}^{2} + 6\,{x_{2}}
^{2}\,{x_{5}}\,{x_{6}}^{2} - 2\,{x_{1}}^{2}\,{x_{5}}^{3} - 2\,{x
_{2}}^{2}\,{x_{5}}^{3} + {x_{1}}^{2}\,{x_{4}}^{2} \\
\mbox{} + {x_{2}}^{2}\,{x_{4}}^{2} - 18\,{x_{1}}^{2} - 18\,{x_{2}
}^{2} + 6\,{x_{1}}\,{x_{6}}\,{x_{8}} - 24\,{x_{1}}\,{x_{2}}^{2}
 + 8\,{x_{1}}^{3} + {x_{1}}^{4}\,{x_{5}}^{2} - 2\,{x_{1}}^{2}\,{x
_{2}}^{2} - {x_{1}}^{4} \\
\mbox{} - {x_{2}}^{4} $

\bigskip

\noindent
$I_2^R(\xxfull)=\\
\mbox{} ~~~\;27 - 2\,{x_{1}}\,{x_{3}}^{2}\,{x_{5}}^{2} + 4\,{x_{2}}\,{x_{3}}^{
2}\,{x_{5}}\,{x_{6}} - 4\,{x_{2}}\,{x_{4}}^{2}\,{x_{5}}\,{x_{6}}
 + 2\,{x_{1}}\,{x_{4}}^{2}\,{x_{5}}^{2} + 2\,{x_{1}}\,{x_{3}}^{2}
\,{x_{6}}^{2} - 2\,{x_{1}}\,{x_{4}}^{2}\,{x_{6}}^{2} \\
\mbox{} + 2\,{x_{1}}^{2}\,{x_{3}}\,{x_{5}}\,{x_{6}}^{2} - 2\,{x_{
2}}^{2}\,{x_{3}}\,{x_{5}}\,{x_{6}}^{2} - 4\,{x_{1}}\,{x_{2}}\,{x
_{3}}\,{x_{6}}^{3} + 2\,{x_{1}}^{2}\,{x_{4}}\,{x_{5}}^{2}\,{x_{6}
} - 2\,{x_{2}}^{2}\,{x_{4}}\,{x_{5}}^{2}\,{x_{6}} \\
\mbox{} + 4\,{x_{1}}\,{x_{2}}\,{x_{4}}\,{x_{5}}^{3} + 6\,{x_{1}}
\,{x_{2}}^{2}\,{x_{5}}^{3} + 6\,{x_{1}}^{3}\,{x_{5}}\,{x_{6}}^{2}
 - 6\,{x_{2}}^{3}\,{x_{5}}^{2}\,{x_{6}} - 6\,{x_{1}}^{2}\,{x_{2}}
\,{x_{6}}^{3} - 2\,{x_{1}}^{3}\,{x_{5}}^{3} \\
\mbox{} + 2\,{x_{2}}^{3}\,{x_{6}}^{3} - 18\,{x_{1}}\,{x_{2}}^{2}
\,{x_{5}}\,{x_{6}}^{2} + 18\,{x_{1}}^{2}\,{x_{2}}\,{x_{5}}^{2}\,{
x_{6}} - 4\,{x_{1}}\,{x_{2}}\,{x_{3}}\,{x_{5}}^{2}\,{x_{6}} + 4\,
{x_{1}}\,{x_{2}}\,{x_{4}}\,{x_{5}}\,{x_{6}}^{2} \\
\mbox{} + 2\,{x_{1}}^{2}\,{x_{3}}\,{x_{5}}^{3} - 2\,{x_{2}}^{2}\,
{x_{3}}\,{x_{5}}^{3} + 2\,{x_{1}}^{2}\,{x_{4}}\,{x_{6}}^{3} - 2\,
{x_{2}}^{2}\,{x_{4}}\,{x_{6}}^{3} - 9\,{x_{8}}^{2} - 9\,{x_{7}}^{
2} - 18\,{x_{3}}^{2} \\
\mbox{} - 18\,{x_{4}}^{2} - 12\,{x_{2}}\,{x_{6}}\,{x_{7}} + 2\,{x
_{7}}^{3} - 6\,{x_{7}}\,{x_{8}}^{2} - 9\,{x_{6}}^{2} - 9\,{x_{5}}
^{2} + 12\,{x_{1}}\,{x_{5}}\,{x_{7}} + 12\,{x_{2}}\,{x_{5}}\,{x_{
8}} \\
\mbox{} - 16\,{x_{1}}\,{x_{2}}\,{x_{3}}\,{x_{6}} + 16\,{x_{1}}\,{
x_{2}}\,{x_{4}}\,{x_{5}} + 2\,{x_{5}}^{3} - 6\,{x_{5}}\,{x_{6}}^{
2} - 2\,{x_{3}}^{2}\,{x_{4}}^{2} + 8\,{x_{1}}^{2}\,{x_{3}}\,{x_{5
}} \\
\mbox{} - 8\,{x_{2}}^{2}\,{x_{3}}\,{x_{5}} + 6\,{x_{1}}\,{x_{3}}
\,{x_{5}} + 6\,{x_{2}}\,{x_{3}}\,{x_{6}} - 6\,{x_{2}}\,{x_{4}}\,{
x_{5}} + 8\,{x_{1}}^{2}\,{x_{4}}\,{x_{6}} - 8\,{x_{2}}^{2}\,{x_{4
}}\,{x_{6}} + 8\,{x_{3}}^{3} \\
\mbox{} - 24\,{x_{3}}\,{x_{4}}^{2} - 16\,{x_{2}}\,{x_{3}}\,{x_{5}
}\,{x_{6}} - 16\,{x_{1}}\,{x_{4}}\,{x_{5}}\,{x_{6}} - 12\,{x_{1}}
^{2}\,{x_{5}}\,{x_{6}}\,{x_{8}} + 12\,{x_{2}}^{2}\,{x_{5}}\,{x_{6
}}\,{x_{8}} \\
\mbox{} + 8\,{x_{1}}\,{x_{4}}^{2}\,{x_{5}} + 4\,{x_{1}}^{2}\,{x_{
3}}\,{x_{6}}\,{x_{8}} + 4\,{x_{2}}^{2}\,{x_{3}}\,{x_{6}}\,{x_{8}}
 + 4\,{x_{1}}^{2}\,{x_{4}}\,{x_{6}}\,{x_{7}} + 4\,{x_{2}}^{2}\,{x
_{4}}\,{x_{6}}\,{x_{7}} \\
\mbox{} + 4\,{x_{1}}^{2}\,{x_{4}}\,{x_{5}}\,{x_{8}} + 4\,{x_{2}}
^{2}\,{x_{4}}\,{x_{5}}\,{x_{8}} - 4\,{x_{1}}\,{x_{3}}\,{x_{5}}^{2
}\,{x_{7}} - 4\,{x_{2}}\,{x_{4}}\,{x_{5}}^{2}\,{x_{7}} + 6\,{x_{1
}}\,{x_{4}}\,{x_{6}} \\
\mbox{} - 8\,{x_{1}}\,{x_{3}}^{2}\,{x_{5}} - 8\,{x_{2}}\,{x_{3}}
^{2}\,{x_{6}} + 8\,{x_{2}}\,{x_{4}}^{2}\,{x_{6}} + 4\,{x_{2}}\,{x
_{3}}\,{x_{5}}^{2}\,{x_{8}} - 4\,{x_{1}}\,{x_{4}}\,{x_{5}}^{2}\,{
x_{8}} + 4\,{x_{2}}\,{x_{3}}\,{x_{6}}^{2}\,{x_{8}} \\
\mbox{} - 4\,{x_{1}}\,{x_{4}}\,{x_{6}}^{2}\,{x_{8}} - 4\,{x_{1}}
^{2}\,{x_{3}}\,{x_{5}}\,{x_{7}} - 4\,{x_{2}}^{2}\,{x_{3}}\,{x_{5}
}\,{x_{7}} + 2\,{x_{3}}\,{x_{5}}^{2}\,{x_{7}} - 2\,{x_{3}}\,{x_{6
}}^{2}\,{x_{7}} \\
\mbox{} + 2\,{x_{1}}\,{x_{3}}\,{x_{4}}^{2}\,{x_{5}} - 2\,{x_{2}}
\,{x_{3}}^{2}\,{x_{4}}\,{x_{5}} + 2\,{x_{1}}\,{x_{3}}^{2}\,{x_{4}
}\,{x_{6}} + 2\,{x_{2}}\,{x_{3}}\,{x_{4}}^{2}\,{x_{6}} - 2\,{x_{2
}}\,{x_{3}}\,{x_{5}}^{2}\,{x_{6}} \\
\mbox{} - 2\,{x_{1}}\,{x_{3}}\,{x_{5}}\,{x_{6}}^{2} - 2\,{x_{1}}
\,{x_{4}}\,{x_{5}}^{2}\,{x_{6}} + 2\,{x_{2}}\,{x_{4}}\,{x_{5}}\,{
x_{6}}^{2} + 4\,{x_{4}}\,{x_{5}}\,{x_{6}}\,{x_{7}} + 4\,{x_{3}}\,
{x_{5}}\,{x_{6}}\,{x_{8}} + {x_{3}}^{2}\,{x_{5}}^{2} \\
\mbox{} + {x_{4}}^{2}\,{x_{5}}^{2} + {x_{3}}^{2}\,{x_{6}}^{2} + {
x_{4}}^{2}\,{x_{6}}^{2} + 4\,{x_{1}}\,{x_{3}}^{2}\,{x_{7}} - 4\,{
x_{1}}\,{x_{4}}^{2}\,{x_{7}} - 4\,{x_{2}}\,{x_{3}}^{2}\,{x_{8}}
 + 4\,{x_{2}}\,{x_{4}}^{2}\,{x_{8}} \\
\mbox{} + 2\,{x_{1}}\,{x_{3}}\,{x_{7}}^{2} - 2\,{x_{1}}\,{x_{3}}
\,{x_{8}}^{2} + 2\,{x_{2}}\,{x_{4}}\,{x_{7}}^{2} - 2\,{x_{2}}\,{x
_{4}}\,{x_{8}}^{2} - 4\,{x_{1}}\,{x_{2}}\,{x_{4}}\,{x_{7}} + 4\,{
x_{1}}\,{x_{2}}\,{x_{3}}\,{x_{8}} \\
\mbox{} - 8\,{x_{2}}\,{x_{3}}\,{x_{4}}\,{x_{7}} - 8\,{x_{1}}\,{x
_{3}}\,{x_{4}}\,{x_{8}} - 4\,{x_{1}}\,{x_{4}}\,{x_{7}}\,{x_{8}}
 + 4\,{x_{2}}\,{x_{3}}\,{x_{7}}\,{x_{8}} - 24\,{x_{1}}\,{x_{2}}\,
{x_{5}}\,{x_{6}}\,{x_{7}} \\
\mbox{} + 2\,{x_{1}}^{2}\,{x_{3}}\,{x_{7}} - 2\,{x_{2}}^{2}\,{x_{
3}}\,{x_{7}} - 12\,{x_{1}}\,{x_{3}}\,{x_{7}} - 12\,{x_{2}}\,{x_{4
}}\,{x_{7}} + 12\,{x_{2}}\,{x_{3}}\,{x_{8}} + 2\,{x_{1}}^{2}\,{x
_{4}}\,{x_{8}} \\
\mbox{} - 2\,{x_{2}}^{2}\,{x_{4}}\,{x_{8}} - 12\,{x_{1}}\,{x_{4}}
\,{x_{8}} - {x_{1}}^{2}\,{x_{3}}^{2}\,{x_{5}}^{2} - {x_{2}}^{2}\,
{x_{3}}^{2}\,{x_{5}}^{2} - {x_{1}}^{2}\,{x_{4}}^{2}\,{x_{5}}^{2}
 - {x_{2}}^{2}\,{x_{4}}^{2}\,{x_{5}}^{2} \\
\mbox{} - {x_{1}}^{2}\,{x_{3}}^{2}\,{x_{6}}^{2} - {x_{2}}^{2}\,{x
_{3}}^{2}\,{x_{6}}^{2} - {x_{1}}^{2}\,{x_{4}}^{2}\,{x_{6}}^{2} -
{x_{2}}^{2}\,{x_{4}}^{2}\,{x_{6}}^{2} + 2\,{x_{1}}\,{x_{3}}^{3}\,
{x_{5}} - 2\,{x_{2}}\,{x_{4}}^{3}\,{x_{5}} \\
\mbox{} + 2\,{x_{1}}\,{x_{4}}^{3}\,{x_{6}} + 2\,{x_{2}}\,{x_{3}}
^{3}\,{x_{6}} + 12\,{x_{1}}\,{x_{6}}\,{x_{7}}\,{x_{8}} + 12\,{x_{
2}}\,{x_{5}}\,{x_{7}}\,{x_{8}} - 12\,{x_{1}}\,{x_{2}}\,{x_{5}}^{2
}\,{x_{8}} \\
\mbox{} + 12\,{x_{1}}\,{x_{2}}\,{x_{6}}^{2}\,{x_{8}} + 4\,{x_{3}}
^{2}\,{x_{5}}\,{x_{7}} - 4\,{x_{4}}^{2}\,{x_{5}}\,{x_{7}} - 6\,{x
_{1}}\,{x_{5}}\,{x_{7}}^{2} + 6\,{x_{1}}\,{x_{5}}\,{x_{8}}^{2} +
6\,{x_{2}}\,{x_{6}}\,{x_{7}}^{2} \\
\mbox{} - 6\,{x_{2}}\,{x_{6}}\,{x_{8}}^{2} + 4\,{x_{3}}\,{x_{6}}
\,{x_{7}}\,{x_{8}} + 4\,{x_{4}}\,{x_{5}}\,{x_{7}}\,{x_{8}} + 8\,{
x_{3}}\,{x_{4}}\,{x_{5}}\,{x_{8}} + 8\,{x_{3}}\,{x_{4}}\,{x_{6}}
\,{x_{7}} - 3\,{x_{1}}^{2}\,{x_{5}}^{2} \\
\mbox{} - 3\,{x_{2}}^{2}\,{x_{5}}^{2} - 3\,{x_{1}}^{2}\,{x_{6}}^{
2} - 3\,{x_{2}}^{2}\,{x_{6}}^{2} + 2\,{x_{3}}\,{x_{5}}\,{x_{7}}^{
2} + 8\,{x_{2}}\,{x_{4}}\,{x_{6}}^{2} + 6\,{x_{1}}^{2}\,{x_{5}}^{
2}\,{x_{7}} \\
\mbox{} - 6\,{x_{2}}^{2}\,{x_{5}}^{2}\,{x_{7}} - 6\,{x_{1}}^{2}\,
{x_{6}}^{2}\,{x_{7}} + 6\,{x_{2}}^{2}\,{x_{6}}^{2}\,{x_{7}} - 2\,
{x_{3}}\,{x_{5}}\,{x_{8}}^{2} - 2\,{x_{4}}\,{x_{6}}\,{x_{7}}^{2}
 + 2\,{x_{4}}\,{x_{6}}\,{x_{8}}^{2} \\
\mbox{} - 4\,{x_{3}}^{2}\,{x_{6}}\,{x_{8}} + 4\,{x_{4}}^{2}\,{x_{
6}}\,{x_{8}} + 8\,{x_{1}}\,{x_{3}}\,{x_{5}}^{2} - 8\,{x_{2}}\,{x
_{4}}\,{x_{5}}^{2} - 8\,{x_{1}}\,{x_{3}}\,{x_{6}}^{2} - 4\,{x_{1}
}\,{x_{3}}\,{x_{6}}^{2}\,{x_{7}} \\
\mbox{} - 4\,{x_{2}}\,{x_{4}}\,{x_{6}}^{2}\,{x_{7}} + 16\,{x_{1}}
\,{x_{3}}\,{x_{4}}\,{x_{6}} - 16\,{x_{2}}\,{x_{3}}\,{x_{4}}\,{x_{
5}} - {x_{3}}^{4} - {x_{4}}^{4} + {x_{3}}^{2}\,{x_{8}}^{2} + {x_{
4}}^{2}\,{x_{8}}^{2} \\
\mbox{} + 12\,{x_{3}}\,{x_{6}}\,{x_{8}} + 12\,{x_{4}}\,{x_{5}}\,{
x_{8}} - 2\,{x_{4}}\,{x_{5}}^{2}\,{x_{8}} + 2\,{x_{4}}\,{x_{6}}^{
2}\,{x_{8}} - 12\,{x_{3}}\,{x_{5}}\,{x_{7}} + 12\,{x_{4}}\,{x_{6}
}\,{x_{7}} \\
\mbox{} - 2\,{x_{1}}^{3}\,{x_{4}}\,{x_{6}} - 2\,{x_{1}}\,{x_{3}}
\,{x_{5}}^{3} - 2\,{x_{2}}\,{x_{3}}\,{x_{6}}^{3} + 2\,{x_{2}}\,{x
_{4}}\,{x_{5}}^{3} - 2\,{x_{1}}\,{x_{4}}\,{x_{6}}^{3} - 2\,{x_{1}
}\,{x_{2}}^{2}\,{x_{3}}\,{x_{5}} \\
\mbox{} - 2\,{x_{1}}^{2}\,{x_{2}}\,{x_{3}}\,{x_{6}} + 2\,{x_{1}}
^{2}\,{x_{2}}\,{x_{4}}\,{x_{5}} - 2\,{x_{1}}\,{x_{2}}^{2}\,{x_{4}
}\,{x_{6}} + {x_{3}}^{2}\,{x_{7}}^{2} + {x_{4}}^{2}\,{x_{7}}^{2}
 + {x_{1}}^{2}\,{x_{3}}^{2} + {x_{2}}^{2}\,{x_{3}}^{2} \\
\mbox{} - 2\,{x_{1}}^{3}\,{x_{3}}\,{x_{5}} - 2\,{x_{2}}^{3}\,{x_{
3}}\,{x_{6}} + 2\,{x_{2}}^{3}\,{x_{4}}\,{x_{5}} + {x_{1}}^{2}\,{x
_{4}}^{2} + {x_{2}}^{2}\,{x_{4}}^{2} - 9\,{x_{1}}^{2} - 9\,{x_{2}
}^{2} + 12\,{x_{1}}\,{x_{6}}\,{x_{8}} \\
\mbox{} - 6\,{x_{1}}\,{x_{2}}^{2} + 2\,{x_{1}}^{3} + 2\,{x_{1}}^{
3}\,{x_{3}}\,{x_{5}}^{2} - 2\,{x_{1}}^{3}\,{x_{3}}\,{x_{6}}^{2}
 - 8\,{x_{1}}\,{x_{3}}\,{x_{4}}\,{x_{5}}\,{x_{6}} - 4\,{x_{1}}^{2
}\,{x_{2}}\,{x_{3}}\,{x_{5}}\,{x_{6}} \\
\mbox{} - 4\,{x_{1}}\,{x_{2}}^{2}\,{x_{4}}\,{x_{5}}\,{x_{6}} + 2
\,{x_{1}}^{2}\,{x_{2}}\,{x_{4}}\,{x_{6}}^{2} - 2\,{x_{1}}\,{x_{2}
}^{2}\,{x_{3}}\,{x_{6}}^{2} + 2\,{x_{1}}\,{x_{2}}^{2}\,{x_{3}}\,{
x_{5}}^{2} - 2\,{x_{1}}^{2}\,{x_{2}}\,{x_{4}}\,{x_{5}}^{2} \\
\mbox{} + 8\,{x_{1}}\,{x_{2}}\,{x_{3}}\,{x_{4}}\,{x_{5}} - 2\,{x
_{1}}^{2}\,{x_{3}}^{2}\,{x_{5}} + 2\,{x_{2}}^{2}\,{x_{3}}^{2}\,{x
_{5}} + 2\,{x_{1}}^{2}\,{x_{4}}^{2}\,{x_{5}} - 2\,{x_{2}}^{2}\,{x
_{4}}^{2}\,{x_{5}} \\
\mbox{} - 2\,{x_{2}}^{3}\,{x_{4}}\,{x_{5}}^{2} + 2\,{x_{2}}^{3}\,
{x_{4}}\,{x_{6}}^{2} - 4\,{x_{2}}^{3}\,{x_{3}}\,{x_{5}}\,{x_{6}}
 - 4\,{x_{1}}^{3}\,{x_{4}}\,{x_{5}}\,{x_{6}} - 4\,{x_{2}}\,{x_{3}
}\,{x_{4}}\,{x_{5}}^{2} \\
\mbox{} + 4\,{x_{2}}\,{x_{3}}\,{x_{4}}\,{x_{6}}^{2} + 4\,{x_{1}}
\,{x_{2}}\,{x_{3}}^{2}\,{x_{6}} - 4\,{x_{1}}\,{x_{2}}\,{x_{4}}^{2
}\,{x_{6}} + 4\,{x_{1}}^{2}\,{x_{3}}\,{x_{4}}\,{x_{6}} - 4\,{x_{2
}}^{2}\,{x_{3}}\,{x_{4}}\,{x_{6}} $

\bigskip

\end{small}

The polynomials $I_1^R$ and $I_2^R$ are both inhomogeneous
polynomials of total degree 6.
$I_1^R$ is a polynomial of degrees $4, 4, 3, 2, 3,
2, 3, 2$ and $I_2^R$ is of
degrees $3, 3, 4, 4, 3, 3, 3, 2$ in variables $\xx$ respectively.

The non-reduced polynomials $I_1$ and $I_2$ are both inhomogeneous
real polynomial of degree 8. $I_1$ is  of degree $4$ in every
variable $\xx$ and $I_2$ is of degree $4$ in the variables
$x_1,x_2,x_5,x_6,x_7,x_8$ and of degree $3$ in the variables
$x_3,x_4$.

%%%%%%%%%%%%%%%%%%%%%%%%%%%%%%%%%%%%%%%%%%%%%%%%%%%%%%%%%%%%%%%%%%%%%%%%%%%%%%%%

\end{document}